\documentclass[aps,prx,twocolumn,longbibliography]{revtex4-1}

\usepackage{graphicx}
\usepackage{bm}
\usepackage{amsmath}
\DeclareMathOperator{\Tr}{Tr}

\begin{document}

\title{Interpretation of saddle-splay and the Oseen-Frank free energy in liquid crystals}
\author{Jonathan V. Selinger}
\affiliation{Department of Physics, Advanced Materials and Liquid Crystal Institute, Kent State University, Kent, Ohio 44242, USA}

\date{January 17, 2019}

\begin{abstract}
This article re-examines a classic question in liquid-crystal physics:  What are the elastic modes of a nematic liquid crystal?  The analysis uses a recent mathematical construction, which breaks the director gradient tensor into four distinct types of mathematical objects, representing splay, twist, bend, and a fourth deformation mode.  With this construction, the Oseen-Frank free energy can be written as the sum of squares of the four modes, and saddle-splay can be regarded as bulk rather than surface elasticity.  This interpretation leads to an alternative way to think about several previous results in liquid-crystal physics, including:  (1)~free energy balance between cholesteric and blue phases, (2)~director deformations in hybrid-aligned-nematic cells, (3)~spontaneous twist of achiral liquid crystals confined in a torus or cylinder, and (4)~curvature of smectic layers.
\end{abstract}

\maketitle

\section{Introduction}

One of the oldest problems in liquid-crystal research is to characterize how the director field can be distorted away from a uniform state, or in other words, to identify the elastic modes of a nematic liquid crystal.  Through the mid-20th century, this issue was investigated in classic work by Oseen~\cite{Oseen1933} and Frank~\cite{Frank1958}, and further by Nehring and Saupe~\cite{Nehring1971,Nehring1972}.  This body of research led to the Oseen-Frank free energy density, which is discussed in many textbooks, such as Ref.~\cite{Kleman2003}, and which is widely used in liquid-crystal science and technology.  This free energy density includes terms representing the cost of three distortion modes---splay, twist, and bend---and also includes a further term called saddle-splay.  The saddle-splay contribution to the free energy density is the total divergence of a vector field.  As a result, the volume integral of this term can be transformed into a surface integral.  For that reason, the saddle-splay contribution is often considered as surface elasticity, in contrast with the splay, twist, and bend contributions, which are bulk elasticity.

Over the years, the role of saddle-splay in liquid-crystal physics has been rather subtle.  In many cases, the saddle-splay free energy is fixed by boundary conditions, and hence does not affect the behavior of a system.  In other cases, the saddle-splay free energy is a variable quantity, and it can induce complex nonuniform structures in the director field~\cite{Crawford1995}.  Many theoretical studies have successfully analyzed the role of saddle-splay in specific systems~\cite{Pairam2013,Koning2014,Davidson2015,Kos2016,Tran2016}, but saddle-splay is still often difficult to understand on any intuitive basis.  This difficulty arises for several reasons, including:  (1)~saddle-splay is normally not visualized by itself, (2)~it can be regarded as either a bulk or surface free energy, and (3)~if it is regarded as a surface contribution, it can accumulate along defects as internal surfaces.

The purpose of this article is to discuss an alternative interpretation of liquid-crystal elasticity, which may help to clarify the role of saddle-splay in the Oseen-Frank free energy. This interpretation is based on a mathematical construction that was recently suggested by Machon and Alexander~\cite{Machon2016}. Their paper decomposes the director gradient tensor into four modes: splay, twist, bend, and a fourth mode that they call $\bm{\Delta}$.  The $\bm{\Delta}$ mode is related to saddle-splay but is not exactly the same---in the following sections, we discuss the terminology and suggest the name ``biaxial splay.''  The Machon-Alexander paper uses this mathematical construction to analyze the topological properties of umbilic lines where $\bm{\Delta}=0$.  Here, we use the same construction for a different purpose. We re-express the Oseen-Frank free energy in terms of the four modes, and use this new expression to re-analyze several previous problems in liquid-crystal physics where saddle-splay was found to be important.  Through this re-analysis, we suggest that the new construction provides a simpler and more intuitive way to understand the role of saddle-splay.

We emphasize that our re-analysis does not change any predictions for experiments. It gives exactly the same predictions as previous studies, because the theories are mathematically equivalent. Hence, the significance of our argument is purely a matter of theoretical understanding.

The plan of this paper is as follows.  In Sec.~II, we explain how the director gradient tensor can be decomposed into the four modes splay, twist, bend, and $\bm{\Delta}$. We first present the argument mathematically, and then visualize and discuss each of the modes.  In Sec.~III, we express the Oseen-Frank free energy in terms of these modes, and show that it takes a simple form as the sum of squares.  In particular, we see how the saddle-splay term is related to splay, twist, and $\bm{\Delta}$.  This relation leads to a discussion of terminology.  In Sec.~IV, we discuss the distinction between double splay and single splay, as well as double twist and cholesteric single twist. This analysis implies that single splay should really be understood as a combination of double splay and $\bm{\Delta}$, and cholesteric single twist as a combination of double twist and $\bm{\Delta}$.  Based on that argument, we assess the free energy balance between cholesteric and blue phases.  In Sec.~V, we apply this analysis to several further examples, particularly director deformations in hybrid-aligned-nematic cells, spontaneous twist of achiral liquid crystals confined in a torus or cylinder, and curvature of smectic layers. In all of these cases, the new analysis provides an alternative way to think about previous results about saddle-splay.  Finally, in Sec.~VI, we discuss some related issues for future research.

We also provide two appendices with specific calculations that might be useful for other investigators. Appendix A shows how the four modes can be expressed in terms of the nematic order tensor $\bm{Q}$. These expressions might be used for analyzing simulations of blue phases, for example. Appendix B shows the two-dimensional (2D) version of how to decompose the director gradient tensor.  In 2D, the only two normal modes are splay and bend.

\section{Director gradient modes}

Following Machon and Alexander~\cite{Machon2016}, we separate the director gradient tensor into four normal modes.  We first explain the decomposition mathematically, and then visualize and discuss each of the modes.

\subsection{Mathematics}

A nematic liquid crystal has a director field $\hat{\bm{n}}(\bm{r})$, and hence a tensor of director gradients $\partial_i n_j$.  Let us first consider the number of degrees of freedom in this tensor.  In three dimensions (3D), the tensor has $3\times3=9$ components.  However, because $\hat{\bm{n}}$ is a unit vector, we must have
\begin{equation}
(\partial_i n_j)n_j = \frac{1}{2}\partial_i(n_j n_j) = \frac{1}{2}\partial_i(1) = 0.
\end{equation}
That equation is actually 3 constraints, for $i=1$ to $3$.  Hence, the tensor $\partial_i n_j$ should have $9-3=6$ degrees of freedom.  In other words, the first leg of $\partial_i n_j$ might have components parallel or penpendicular to $\hat{\bm{n}}$, but the second leg must be perpendicular to $\hat{\bm{n}}$.

We can break $\partial_i n_j$ into parts where the first leg is parallel or perpendicular to $\hat{\bm{n}}$,
\begin{equation}
\partial_i n_j = -n_i B_j + \alpha_{ij},
\label{decomposition1}
\end{equation}
where $\bm{B}$ is a vector perpendicular to $\hat{\bm{n}}$ and $\alpha_{ij}$ is a tensor in the plane perpendicular to $\hat{\bm{n}}$.  Contracting both sides of Eq.~(\ref{decomposition1}) with $n_i$ gives
\begin{equation}
\bm{B} = -(\hat{\bm{n}}\cdot\bm{\nabla})\hat{\bm{n}} = \hat{\bm{n}}\times(\bm{\nabla}\times\hat{\bm{n}}).
\label{benddefinition}
\end{equation}
Hence, $\bm{B}$ is the standard \emph{bend vector} in liquid-crystal physics.  Because $\bm{B}$ is perpendicular to $\hat{\bm{n}}$, it has two degrees of freedom.

Now we are left with the tensor $\alpha_{ij}$ in the plane perpendicular to $\hat{\bm{n}}$, which has 4 degrees of freedom.  We can break it into an antisymmetric tensor $\beta_{ij}$ and symmetric tensor $\gamma_{ij}$.  Because $\beta_{ij}$ is an antisymmetric tensor in the plane perpendicular to $\hat{\bm{n}}$, it can be expressed as $\beta_{ij}=\frac{1}{2}T\epsilon_{ijk}n_k$, for some pseudoscalar $T$.  Hence, the director gradient tensor becomes
\begin{equation}
\partial_i n_j = -n_i B_j + \frac{1}{2}T\epsilon_{ijk}n_k + \gamma_{ij}.
\label{decomposition2}
\end{equation}
Contracting both sides of Eq.~(\ref{decomposition2}) with $\epsilon_{ijm}n_m$ gives
\begin{equation}
T=\hat{\bm{n}}\cdot(\bm{\nabla}\times\hat{\bm{n}}).
\label{twistdefinition}
\end{equation}
Hence, $T$ is the standard \emph{twist} in liquid-crystal physics.  Because $T$ is a pseudoscalar, it has one degree of freedom.

At this point, we are left with the symmetric tensor $\gamma_{ij}$ in the plane perpendicular to $\hat{\bm{n}}$, which has 3 degrees of freedom.  We can break it into its trace $S$, multiplied by half of the identity tensor in the plane perpendicular to $\hat{\bm{n}}$, which is $\frac{1}{2}(\delta_{ij}-n_i n_j)$, plus a traceless symmetric tensor $\Delta_{ij}$.  Hence, the director gradient tensor becomes
\begin{equation}
\partial_i n_j = -n_i B_j + \frac{1}{2}T\epsilon_{ijk}n_k + \frac{1}{2}S(\delta_{ij}-n_i n_j) + \Delta_{ij}.
\label{decomposition}
\end{equation}
Taking the trace of Eq.~(\ref{decomposition}) gives
\begin{equation}
S=\bm{\nabla}\cdot\hat{\bm{n}}.
\label{splaydefinition}
\end{equation}
Hence, $S$ is the standard \emph{splay} in liquid-crystal physics.  Because $S$ is a scalar, it has one degree of freedom.

Finally, we have the traceless symmetric tensor $\Delta_{ij}$ in the plane perpendicular to $\hat{\bm{n}}$.  As a traceless symmetric tensor in a plane, it has two degrees of freedom.  For an explicit expression for $\Delta_{ij}$, we combine Eq.~(\ref{decomposition}) with its transpose to obtain
\begin{align}
\Delta_{ij} = & \frac{1}{2} [\partial_i n_j + \partial_j n_i + n_i B_j + n_j B_i - S (\delta_{ij} - n_i n_j)] \nonumber\\
=& \frac{1}{2} [\partial_i n_j + \partial_j n_i - n_i n_k \partial_k n_j - n_j n_k \partial_k n_i \nonumber\\
&\quad - \delta_{ij} \partial_k n_k + n_i n_j \partial_k n_k ].
\label{biaxialsplaydefinition}
\end{align}
We note that $\Delta_{ij}=\Delta_{ji}$, $\Delta_{ii}=0$, $n_i \Delta_{ij}=0$, and $\Delta_{ij}n_j =0$.

Equation~(\ref{decomposition}) decomposes the director gradient tensor $\partial_i n_j$ into the four normal modes $\bm{B}$, $T$, $S$, and $\bm{\Delta}$.  Together, these modes account for the six degrees of freedom in $\partial_i n_j$.  These modes are four distinct types of mathematical objects:  $\bm{B}$ is a vector, $T$ a pseudoscalar, $S$ a scalar, and $\bm{\Delta}$ a symmetric traceless tensor.  In a more precise mathematical language, Machon and Alexander~\cite{Machon2016} write that these modes are distinct irreducible representations of the rotation group.

\subsection{Visualization and discussion of each mode}

In this section, we discuss each of the four modes by visualizing director configurations in which all the other modes are zero, at least locally, if not globally.

\subsubsection{Bend $\bm{B}$}

\begin{figure*}
\begin{tabular}{ccc}
(a) Bend $\bm{B}$ & (b) Bend $\bm{B}$ & (c) Twist $T$ \\
\includegraphics[height=4.4cm]{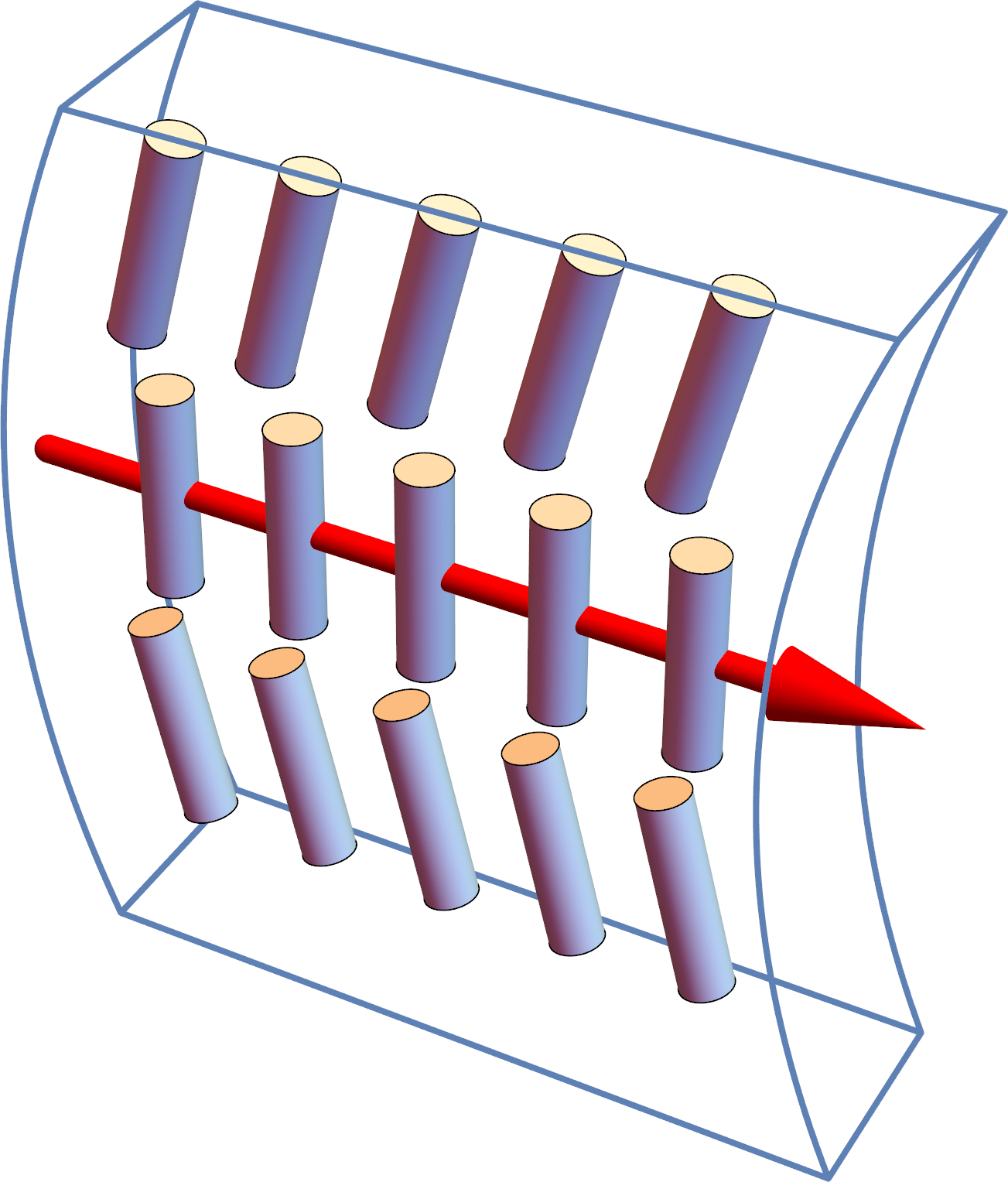} & \includegraphics[height=4.4cm]{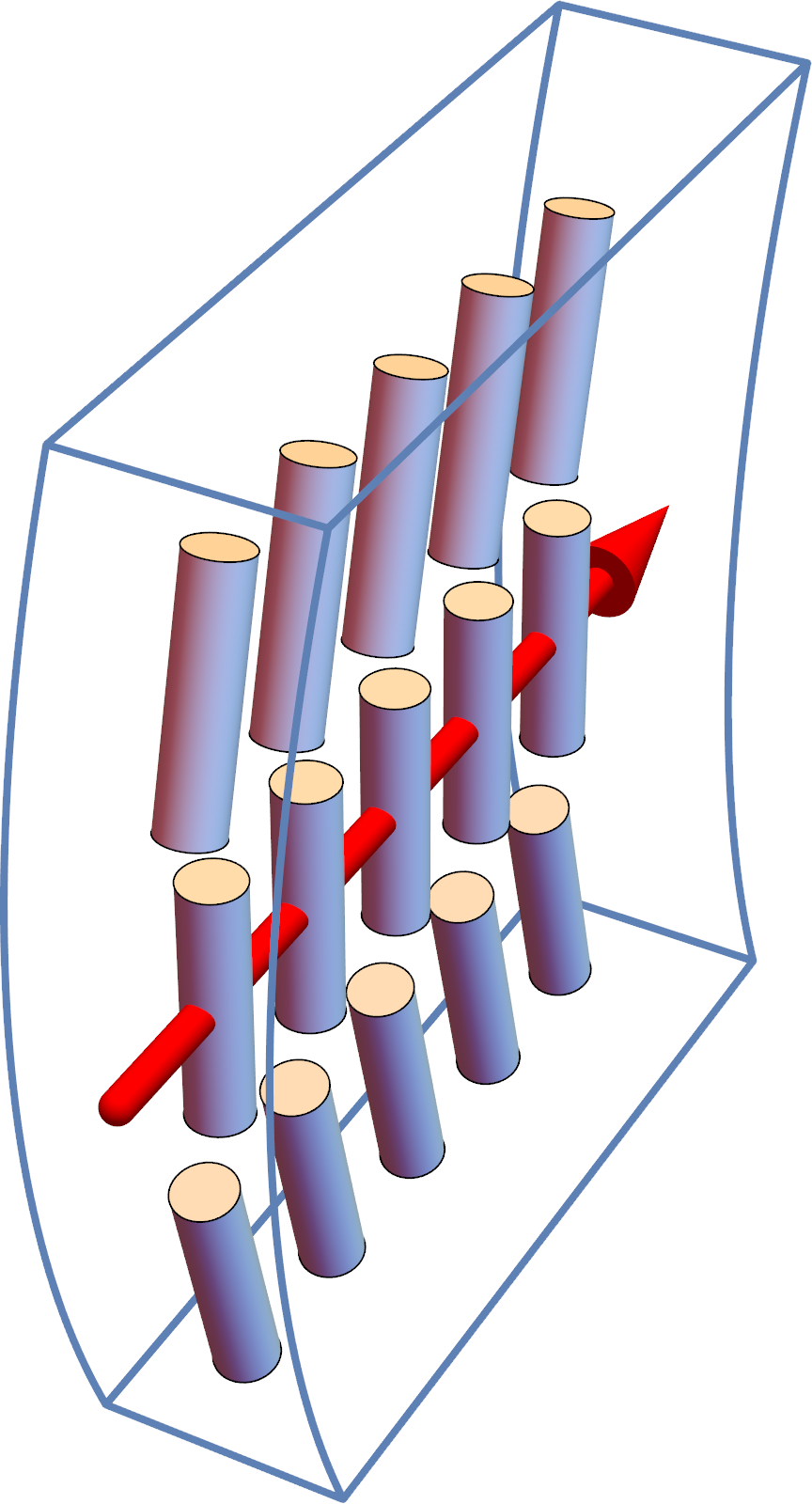} & \includegraphics[height=4.4cm]{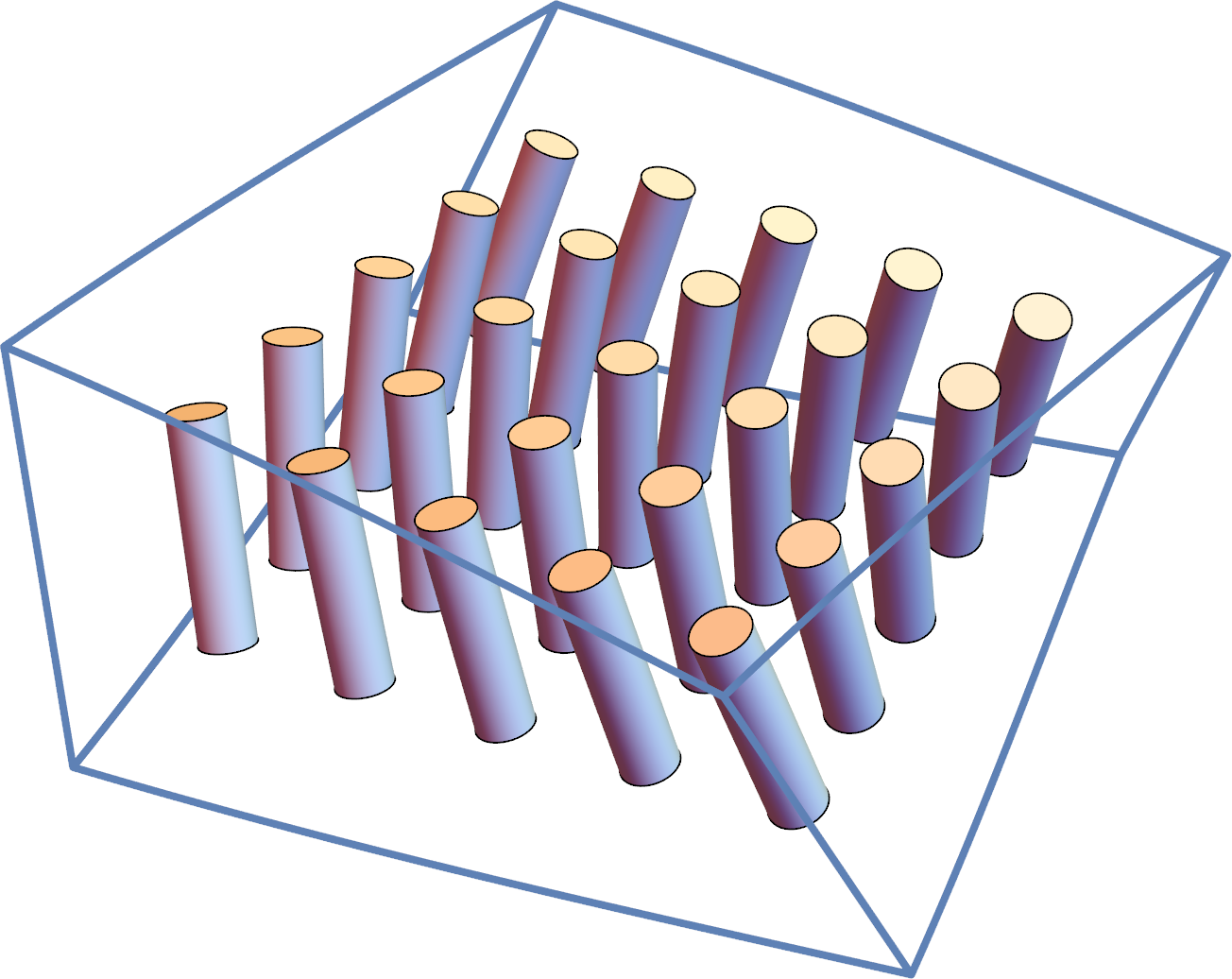} \\
(d) Splay $S$ & (e) Biaxial splay $\Delta_{ij}$ & (f) Biaxial splay $\Delta_{ij}$\\
\includegraphics[height=4.4cm]{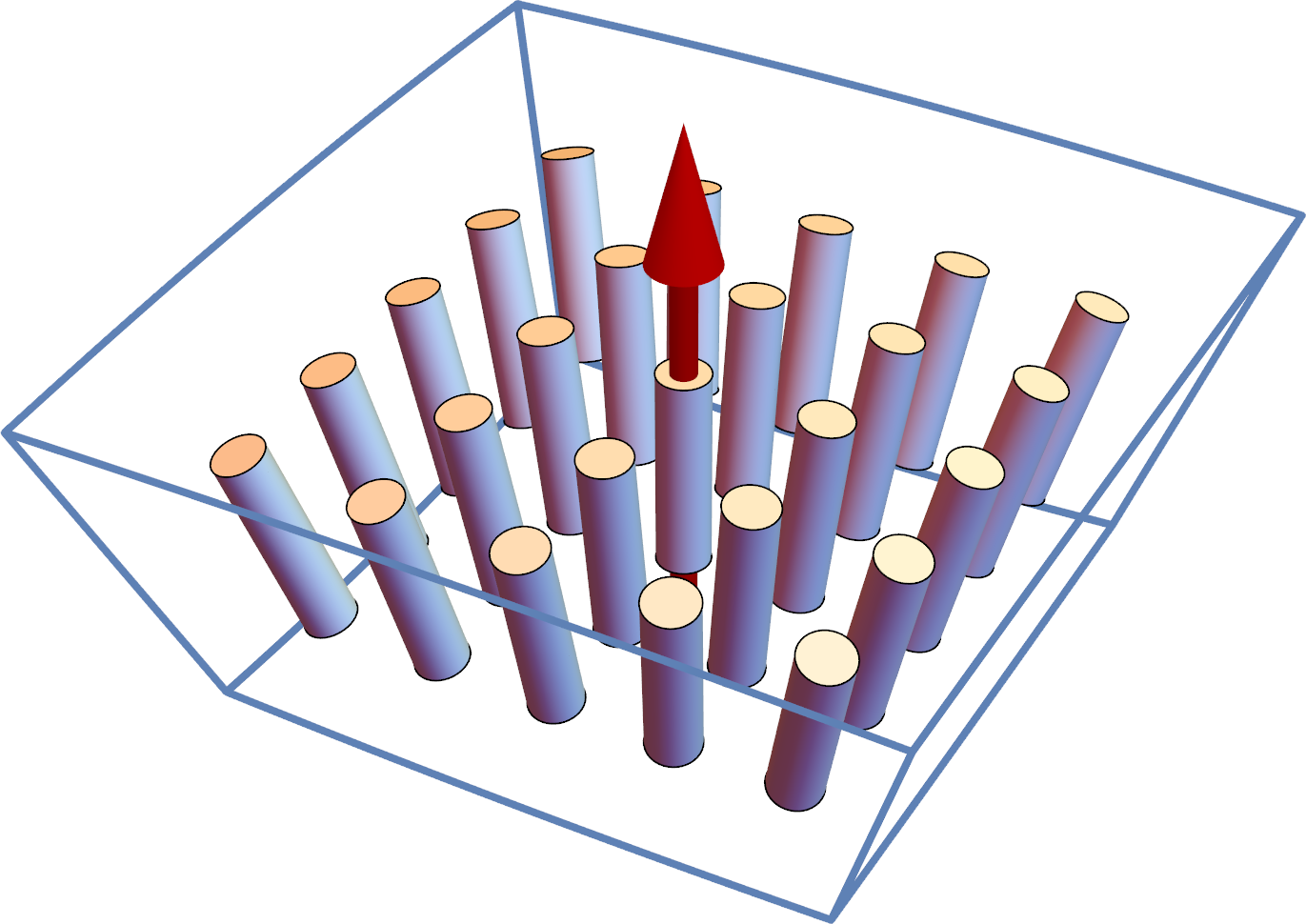} & \includegraphics[height=4.4cm]{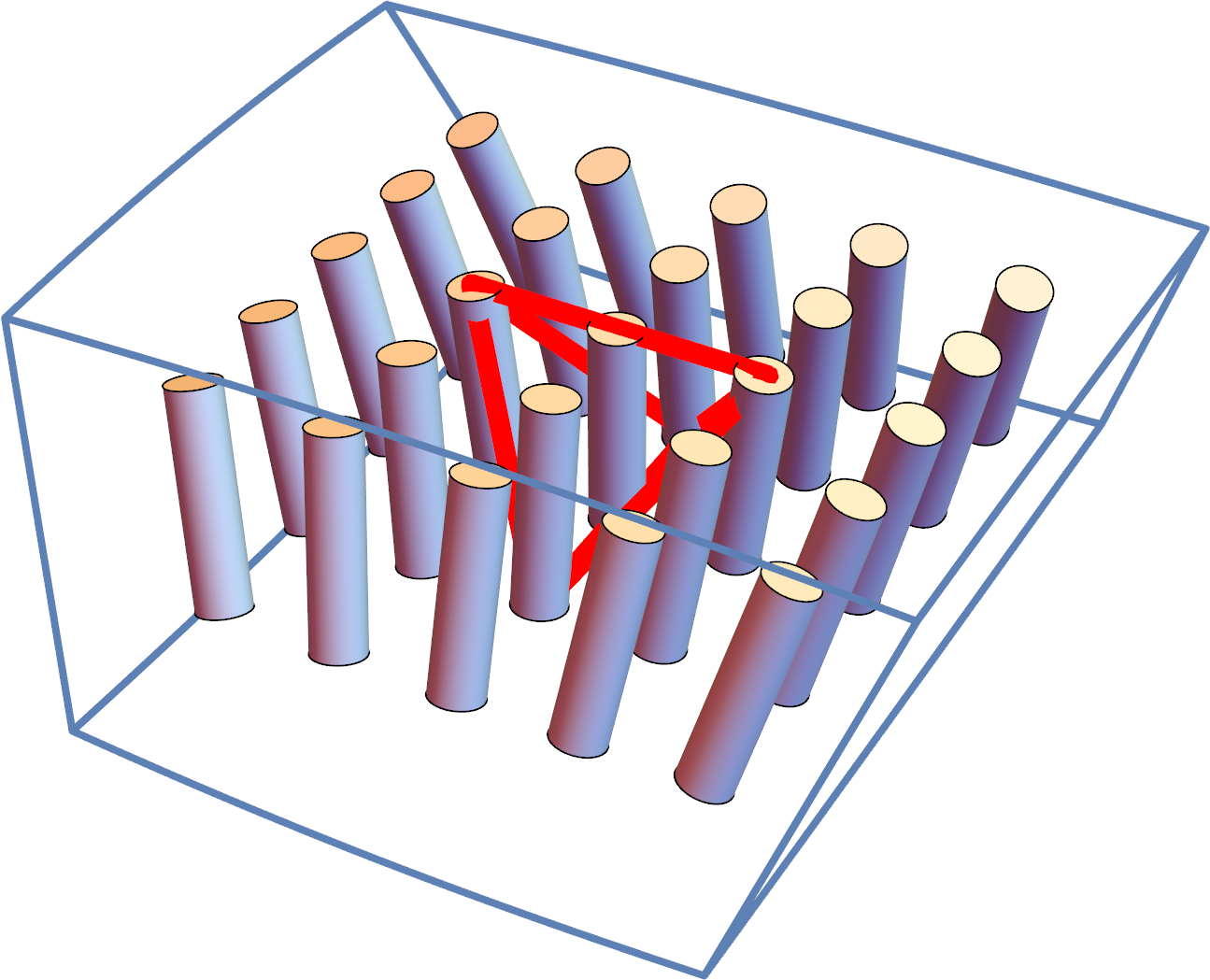} & \includegraphics[height=4.4cm]{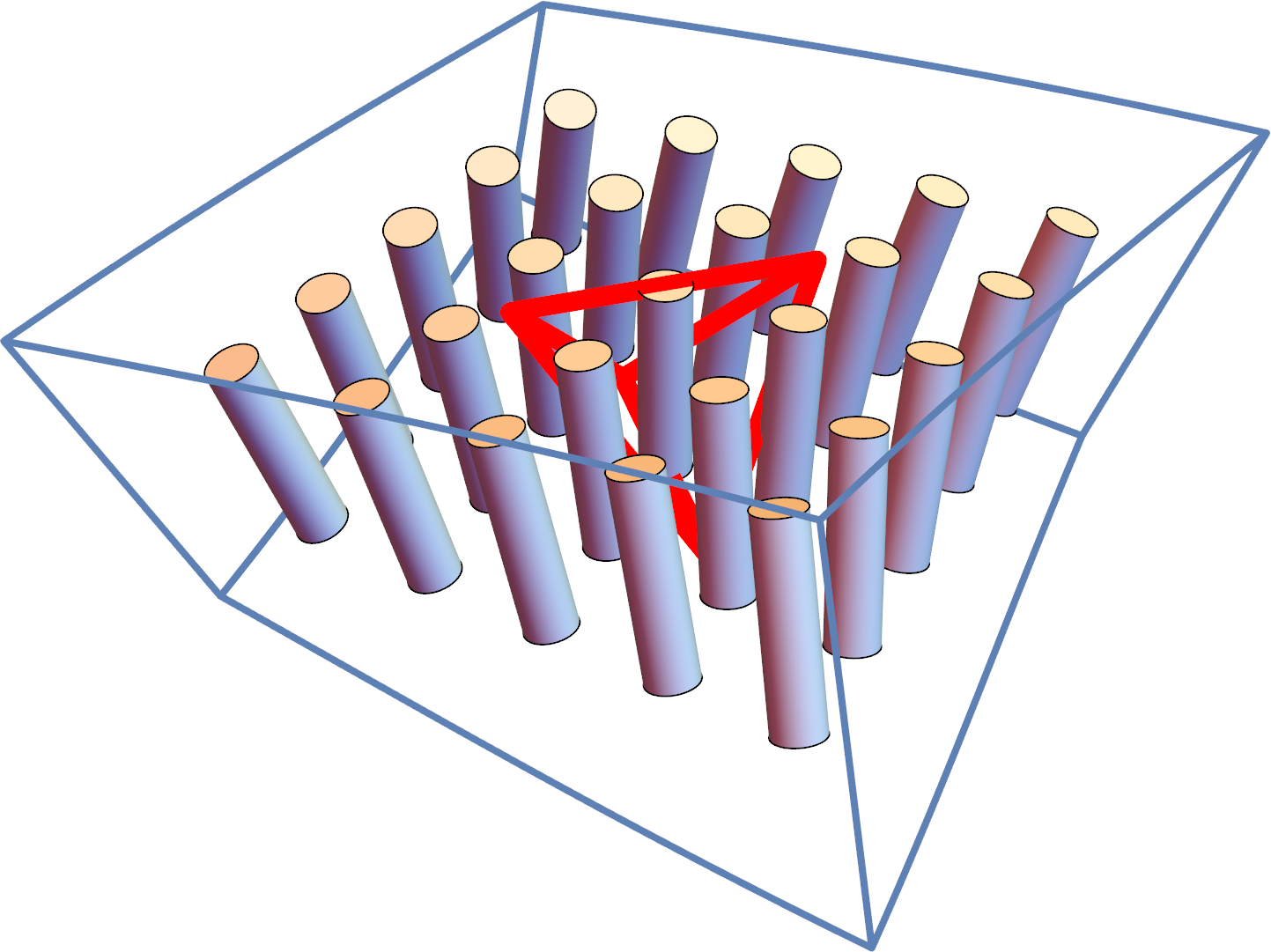}
\end{tabular}
\caption{(Color online) Schematic illustrations of the four director deformation modes.  (a,b)~Two components of bend, with the red arrows representing the bend vector $\bm{B}$.  (c)~Twist (i.e. double twist).  (d)~Splay (i.e. double splay), with the red arrow representing the splay vector $\bm{S}=S\hat{\bm{n}}$.  (e,f)~Two components of biaxial splay $\Delta_{ij}$, with the red tetrahedra representing the third-rank tensor $\Delta_{ij}n_k$.}
\end{figure*}

A bend deformation has the structure shown in Fig.~1(a,b).  It has the standard form of bend in liquid-crystal physics, with $\hat{\bm{n}}$ varying along the direction parallel to the local $\hat{\bm{n}}$.  It has two components, because the change in $\hat{\bm{n}}$ can be in either of two directions perpendicular to the local $\hat{\bm{n}}$.  Hence, it can be represented by a vector in the plane perpendicular to the local $\hat{\bm{n}}$, as shown by the red arrows in the figures.

It is possible to fill up 3D Euclidean space with pure bend.  For example, consider the director field
\begin{equation}
\hat{\bm{n}}(x,y,z)=\frac{(-y,x,0)}{\sqrt{x^2+y^2}}.
\end{equation}
Explicit calculations give
\begin{equation}
\bm{B}=\frac{(x,y,0)}{x^2+y^2},\quad
T=0,\quad
S=0,\quad
\bm{\Delta}=0.
\end{equation}
Hence, this example has pure bend, which decreases in magnitude as we move away from the $z$-axis.

In a nematic liquid crystal, $\hat{\bm{n}}$ and $-\hat{\bm{n}}$ are equivalent ways to describe the same state.  The bend vector $\bm{B}$ defined by Eq.~(\ref{benddefinition}) is invariant under the transformation $\hat{\bm{n}}\to-\hat{\bm{n}}$, and hence it is a physical object that does not depend on this arbitrary choice of sign.

\subsubsection{Twist $T$}

A twist deformation has the structure shown in Fig.~1(c).  In the liquid-crystal literature, this type of deformation is commonly known as ``double twist,'' because $\hat{\bm{n}}$ varies in both directions perpendicular to the local $\hat{\bm{n}}$.  We emphasize that twist $T$ means double twist.  In other words, the deformation with nonzero $T$ but zero $\bm{B}$, $S$, and $\bm{\Delta}$ is double twist, not single twist.  This statement must be true because $T$ is a pseudoscalar, and hence it has no direction in the plane perpendicular to $\hat{\bm{n}}$.  If the deformation were single twist, with a helical axis perpendicular to $\hat{\bm{n}}$, then it could not be described by a pseudoscalar.  We will discuss single twist in Sec.~IV.

It is impossible to fill up 3D Euclidean space with pure double twist, but we can construct pure double twist locally.  For example, consider the director field
\begin{equation}
\hat{\bm{n}}(x,y,z)=\frac{(-qy,qx,1)}{\sqrt{1+q^2(x^2+y^2)}}
\end{equation}
for $q\sqrt{x^2+y^2}\ll1$.  Along the $z$-axis, for $x=y=0$, explicit calculations give
\begin{equation}
\bm{B}=0,\quad
T=2q,\quad
S=0,\quad
\bm{\Delta}=0.
\end{equation}
Hence, this example has pure double twist along the $z$-axis.  (Farther from the $z$-axis, it has a mixture of double twist and bend.)

The twist pseudoscalar $T$ defined by Eq.~(\ref{twistdefinition}) is invariant under the transformation $\hat{\bm{n}}\to-\hat{\bm{n}}$, and hence it is a physical object that does not depend on this arbitrary choice of sign.

\subsubsection{Splay $S$}

A splay deformation has the structure shown in Fig.~1(d).  By analogy with double twist, this type of deformation might be called ``double splay,'' because $\hat{\bm{n}}$ varies in both directions perpendicular to the local $\hat{\bm{n}}$.  As in the previous case, we emphasize that splay $S$ means double splay; i.e.\ the deformation with nonzero $S$ but zero $\bm{B}$, $T$, and $\bm{\Delta}$ is double splay, not single splay.  This statement must be true because $S$ is a scalar, and hence has no direction in the plane perpendicular to $\hat{\bm{n}}$.  If the deformation were single splay, with variation in only one direction perpendicular to $\hat{\bm{n}}$, then it could not be described by a scalar.  We will discuss single splay in Sec.~IV.

It is possible to fill up 3D Euclidean space with pure double splay.  For example, consider the director field for a hedgehog,
\begin{equation}
\hat{\bm{n}}(x,y,z)=\frac{(x,y,z)}{\sqrt{x^2+y^2+z^2}}.
\end{equation}
Explicit calculations give
\begin{equation}
\bm{B}=0,\quad
T=0,\quad
S=\frac{2}{\sqrt{x^2+y^2+z^2}},\quad
\bm{\Delta}=0.
\end{equation}
Hence, this example has pure double splay, which decreases in magnitude as we move away from the origin.

The splay scalar $S$ defined by Eq.~(\ref{splaydefinition}) changes sign under the transformation $\hat{\bm{n}}\to-\hat{\bm{n}}$.  If we want a physical object that does not depend on this arbitrary choice of sign, we can construct the splay vector $\bm{S}=S\hat{\bm{n}}={\hat{\bm{n}}(\bm{\nabla}\cdot\hat{\bm{n}})}$.  This splay vector is well-known in the theory of flexoelectricity~\cite{Meyer1969}.  It is shown by the red arrow in Fig.~1(d).

\subsubsection{``Biaxial splay'' $\bm{\Delta}$}

The fourth deformation mode $\bm{\Delta}$ has the structure shown in Fig.~1(e,f).  In this deformation, $\hat{\bm{n}}$ tips outward along one axis perpendicular to the local $\hat{\bm{n}}$, and tips inward along the other axis perpendicular to the local $\hat{\bm{n}}$.  In other words, there is a combination of positive splay along one axis and negative splay along the other axis.  The symmetry of this deformation is similar to a biaxial nematic liquid crystal, because of the two distinct axes perpendicular to $\hat{\bm{n}}$.  For that reason, we suggest that this deformation might be called ``biaxial splay.''  We will discuss the terminology further in Sec.~III.

The biaxial splay deformation has two components, which are rotated with respect to each other by $45^\circ$ in the plane perpendicular to $\hat{\bm{n}}$.  In Fig.~1(e), the splay is outward along $\pm(1,0,0)$ and inward along $\pm(0,1,0)$.  By comparison, in Fig.~1(f), the splay is outward along $\pm(1/\sqrt{2},1/\sqrt{2},0)$ and inward along $\pm(1/\sqrt{2},-1/\sqrt{2},0)$.  A further rotation of $45^\circ$ gives a splay outward along $\pm(0,1,0)$ and inward along $\pm(1,0,0)$, which is just the negative of the first component in Fig.~1(e).

It is impossible to fill up 3D Euclidean space with pure biaxial splay, but we can construct pure biaxial splay locally.  For example, consider the director field
\begin{equation}
\hat{\bm{n}}(x,y,z)=\frac{(qx,-qy,1)}{\sqrt{1+q^2(x^2+y^2)}}
\end{equation}
for $q\sqrt{x^2+y^2}\ll1$.  Along the $z$-axis, for $x=y=0$, explicit calculations give
\begin{equation}
\bm{B}=0,\quad
T=0,\quad
S=0,\quad
\bm{\Delta}=
\begin{pmatrix}
q &  0 & 0\\
0 & -q & 0\\
0 &  0 & 0
\end{pmatrix}
.
\end{equation}
Hence, this example has pure biaxial splay along the $z$-axis.  (Farther from the $z$-axis, it has a mixture of biaxial splay with bend and double splay.)

The biaxial splay tensor $\Delta_{ij}$ defined by Eq.~(\ref{biaxialsplaydefinition}) changes sign under the transformation $\hat{\bm{n}}\to-\hat{\bm{n}}$.  If we want a physical object that does not depend on this arbitrary choice of sign, we can construct the third-rank tensor $\Delta_{ij}n_k$.  This third-rank tensor is represented by the red tetrahedra in Fig.~1(e,f).

\section{Free energy, saddle-splay, and $K_{24}$}

The Oseen-Frank free energy gives the elastic free energy associated with deformations in the director field.  We consider first the simplified version of the free energy with equal elastic constants, and then the full free energy with unequal elastic constants.

In the simple approximation of equal elastic constants, the Oseen-Frank free energy density is
\begin{equation}
F=\frac{1}{2}K(\partial_i n_j)(\partial_i n_j).
\label{oseenfranksimple}
\end{equation}
By inserting Eq.~(\ref{decomposition}) for $\partial_i n_j$ into Eq.~(\ref{oseenfranksimple}), we obtain
\begin{equation}
F=\frac{1}{4}KS^2 + \frac{1}{4}KT^2 + \frac{1}{2}K|\bm{B}|^2 + \frac{1}{2}K\Tr(\bm{\Delta}^2),
\end{equation}
where $\Tr(\bm{\Delta}^2)=\Delta_{ij}\Delta_{ji}$.  This expression shows that all four of the modes cost elastic free energy.  There are no cross terms between the modes.  Indeed, bilinear cross terms are forbidden by symmetry:  There is no bilinear coupling that can generate a scalar for the free energy density.  (The coupling $S T$ is a pseudoscalar, which is permitted in a chiral liquid crystal, but not in an achiral nematic phase.)  In this approximation, the coefficients of $S^2$, $T^2$, $|\bm{B}|^2$, and $\Tr(\bm{\Delta}^2)$ are all the same, except for the factors of $\frac{1}{4}$ and $\frac{1}{2}$.  To understand these factors, we might say that $S$ and $T$ are double deformations, while $\bm{B}$ and $\bm{\Delta}$ are single deformations with two components.

In general, the full Oseen-Frank free energy density is conventionally written as~\cite{Kleman2003}
\begin{align}
\label{oseenfrankconventional}
F=&\frac{1}{2}K_{11}S^2 + \frac{1}{2}K_{22}T^2 + \frac{1}{2}K_{33}|\bm{B}|^2 \\
&-K_{24}\bm{\nabla}\cdot\left[\hat{\bm{n}}(\bm{\nabla}\cdot\hat{\bm{n}})+\hat{\bm{n}}\times(\bm{\nabla}\times\hat{\bm{n}})\right],\nonumber
\end{align}
where the last term is the saddle-splay term.  In the literature, there are some variations in the notation for the saddle-splay term.  Instead of $K_{24}$, the coefficient is sometimes written as $\frac{1}{2}K_{24}$ or as $(K_{22}+K_{24})$.  Those variations are not important for the following argument, which still applies with a minor change of notation.

The saddle-splay term can be expressed in terms of the four modes discussed in the previous section.  An explicit calculation gives
\begin{align}
\label{saddlesplay}
&\bm{\nabla}\cdot\left[\hat{\bm{n}}(\bm{\nabla}\cdot\hat{\bm{n}})+\hat{\bm{n}}\times(\bm{\nabla}\times\hat{\bm{n}})\right]\\
&=\bm{\nabla}\cdot\left[\hat{\bm{n}}(\bm{\nabla}\cdot\hat{\bm{n}})-(\hat{\bm{n}}\cdot\bm{\nabla})\hat{\bm{n}})\right]
=\partial_j \left[n_j \partial_i n_i - n_i \partial_i n_j \right]\nonumber\\
&=(\partial_i n_i)(\partial_j n_j)-(\partial_i n_j)(\partial_j n_i)
=\frac{1}{2}S^2 +\frac{1}{2}T^2 - \Tr(\bm{\Delta}^2)\nonumber
\end{align}
Combining Eqs.~(\ref{oseenfrankconventional}) and~(\ref{saddlesplay}), the Oseen-Frank free energy density becomes
\begin{align}
\label{oseenfranknew}
F=&\frac{1}{2}K_{11}S^2 + \frac{1}{2}K_{22}T^2 + \frac{1}{2}K_{33}|\bm{B}|^2 \nonumber\\
&-K_{24}\left[\frac{1}{2}S^2 +\frac{1}{2}T^2 - \Tr(\bm{\Delta}^2)\right]\nonumber\\
=&\frac{1}{2}(K_{11}-K_{24})S^2 + \frac{1}{2}(K_{22}-K_{24})T^2 \nonumber\\
&+\frac{1}{2}K_{33}|\bm{B}|^2 + K_{24}\Tr(\bm{\Delta}^2) .
\end{align}
This expression shows that all four of the modes cost different amounts of elastic free energy.  The elastic constant for (double) splay is $(K_{11}-K_{24})$, the elastic constant for (double) twist is $(K_{22}-K_{24})$, the elastic constant for bend is $K_{33}$, and the elastic constant for the mode $\bm{\Delta}$ is $2K_{24}$.  The simple approximation of equal elastic constants then corresponds to $K_{11}=K_{22}=K_{33}=2K_{24}\equiv K$.

We can now make a remark about the terminology for the mode $\bm{\Delta}$.  Because of the saddle-like shape of the deformation in Figs.~1(e,f), one might be tempted to consider $\bm{\Delta}$ as saddle-splay.  However, Eq.~(\ref{saddlesplay}) shows that the saddle-splay term in the free energy is actually a combination of $S$, $T$, and $\bm{\Delta}$.  Because ``saddle-splay'' is already well-established as the name for that term in the free energy, we cannot use the same name for $\bm{\Delta}$.  Hence, the mode $\bm{\Delta}$ needs another name.  Machon and Alexander~\cite{Machon2016} refer to this mode as ``anisotropic orthogonal gradients of $\hat{\bm{n}}$,'' but that phrase is too long for common use.  For that reason, we suggest the name ``biaxial splay,'' to emphasize the similarity with the symmetry of a biaxial nematic phase.

Up to now, we have not yet used the fact that the saddle-splay term is a total divergence.  Because it is a total divergence, the volume integral of this term can be reduced to a surface integral.  Hence, from Eq.~(\ref{saddlesplay}), the volume integral of $[\frac{1}{2}S^2 +\frac{1}{2}T^2 - \Tr(\bm{\Delta}^2)]$ can be reduced to a surface integral.  In many liquid-crystal systems, that surface integral is a constant determined by the boundary conditions.  In those systems, liquid-crystal theorists normally use this constraint to eliminate $\bm{\Delta}$ from the theory, and work in terms of the three remaining modes $S$, $T$, and $\bm{B}$.  As an alternative, in principle, one might eliminate $S$ or $T$ from the theory, and work in terms of the other three modes.

We do not actually recommend either of those alternatives.  Rather, we suggest treating all four of the modes as bulk elastic modes, and using the bulk free energy density of Eq.~(\ref{oseenfranknew}).  This expression explicitly shows the free energy associated with each of the four modes, and it is clearly positive-definite as long as the four coefficients are positive.  Moreover, it applies even to liquid-crystal systems in which the surface integral is not a constant determined by the boundary conditions, but rather is variable.

In the following sections, we discuss several problems that have been previously treated using the concept of saddle-splay as a surface integral, and re-analyze them in terms of the four bulk elastic modes.  We argue that this view gives an interesting new perspective on those problems, although it still gives the same predictions for experiments.

\section{Single vs.\ double deformations}

\subsection{Planar (single) splay}

\begin{figure*}
\begin{tabular}{cccc}
(a) Planar (single) splay & (b) Splay Frederiks transition & (c) Cholesteric (single) twist & (d) Twist Frederiks transition \\
\includegraphics[height=4.2cm]{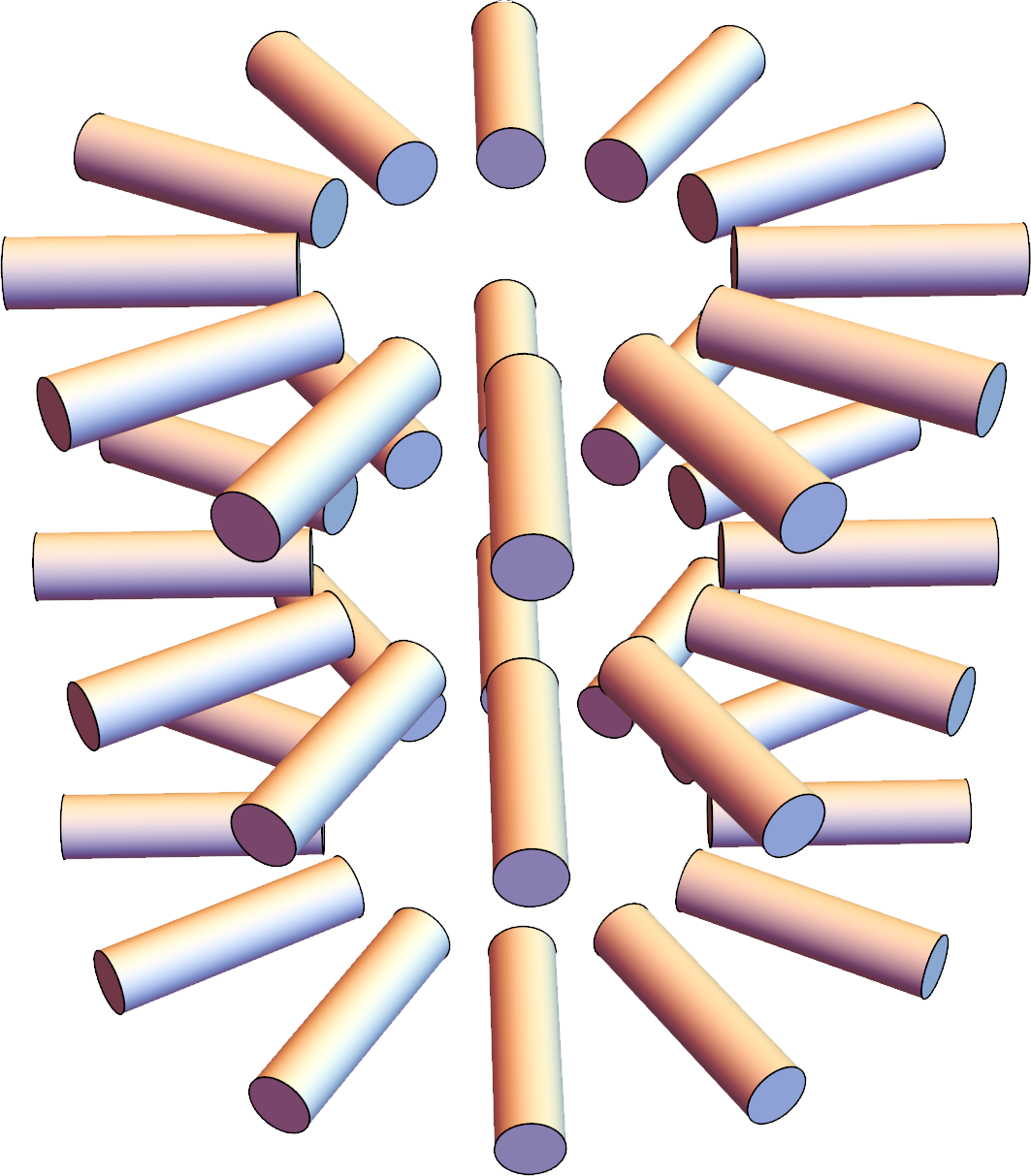} & \includegraphics[height=4.2cm]{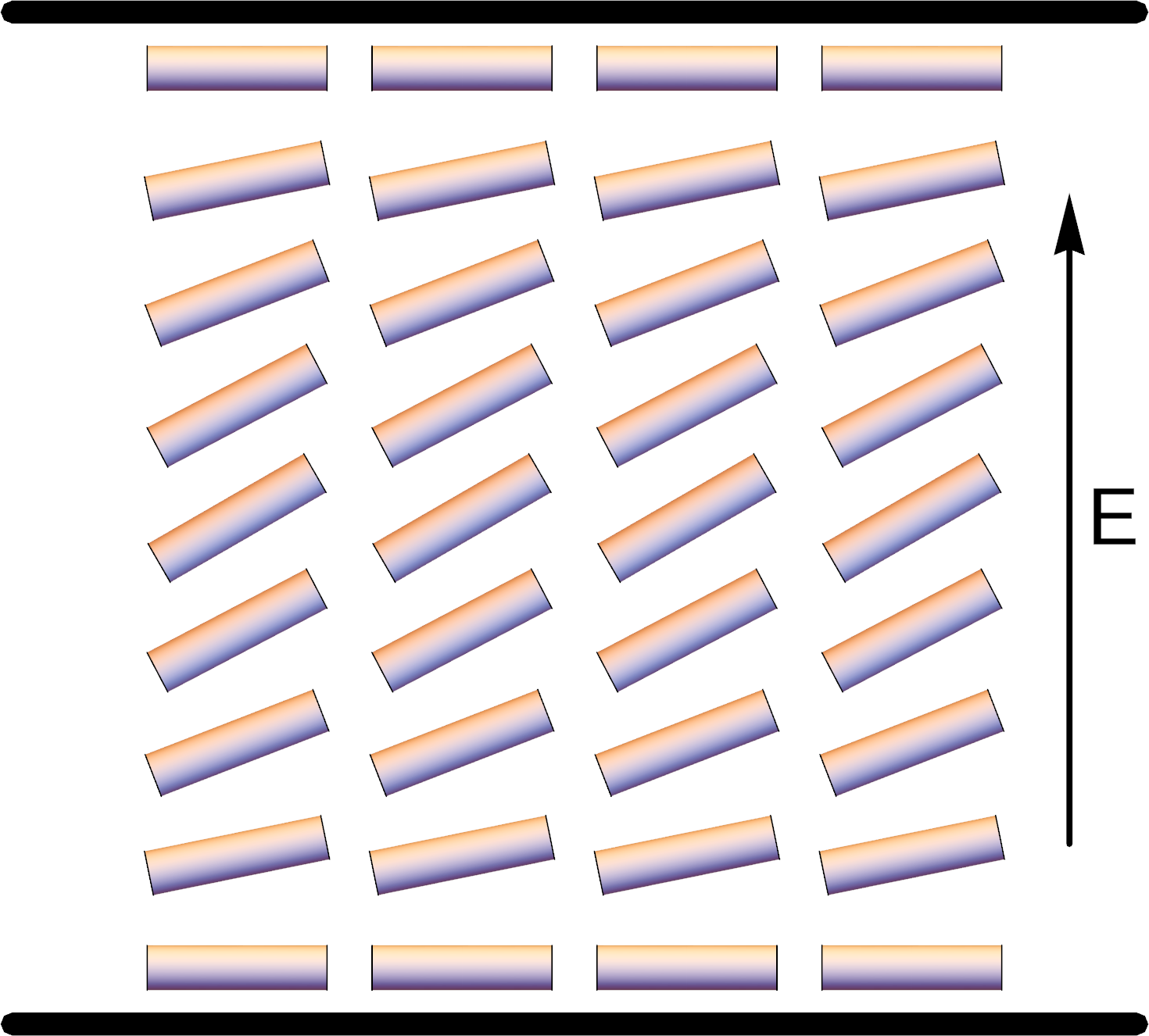} & \includegraphics[height=4.2cm]{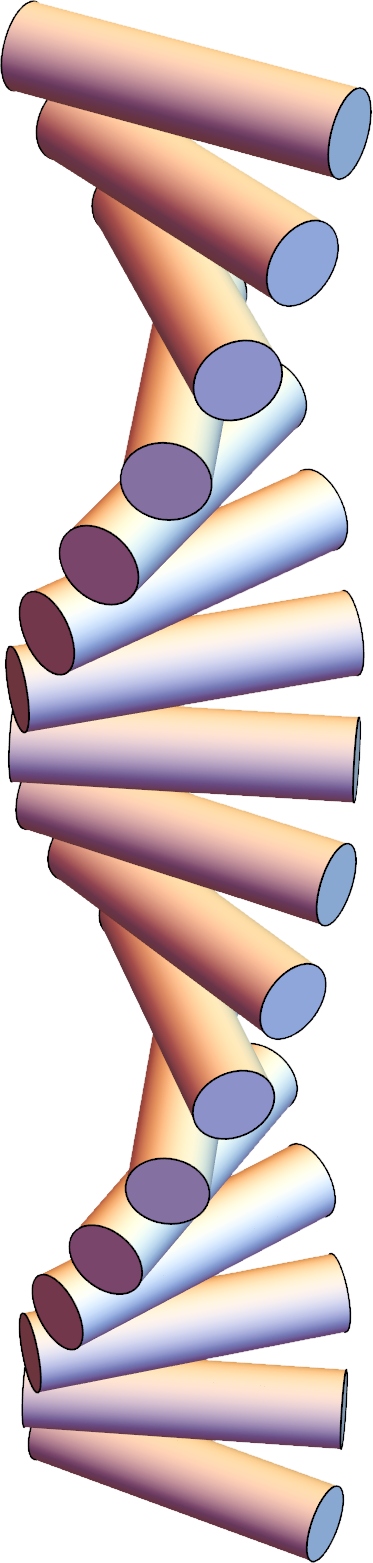} & \includegraphics[height=4.2cm]{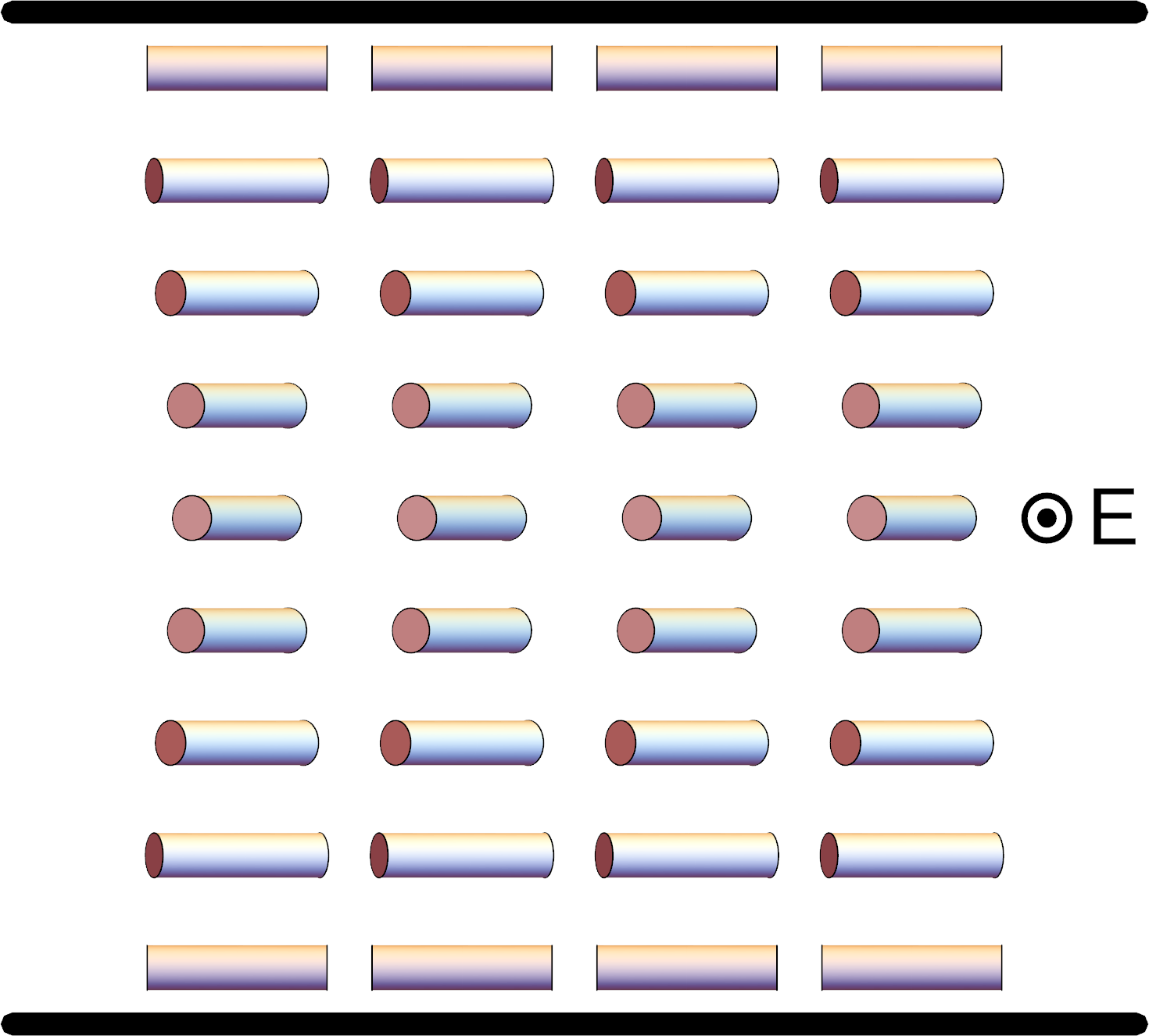} \\
\end{tabular}
\caption{(Color online) Single deformations in the director field.  (a)~Planar or single splay.  (b)~Splay Frederiks transition under an electric field $\bm{E}$, showing that the deformation is single splay.  (c)~Cholesteric or single twist.  (d)~Twist Frederiks transition under an electric field $\bm{E}$, showing that the deformation is single twist.}
\end{figure*}

In Sec.~II, we argued that a deformation with nonzero $S$ but zero $\bm{B}$, $T$, and $\bm{\Delta}$ is 3D double splay, as in a hedgehog.  For comparison, we might consider a deformation with splay only in the 2D plane, as in slices of pizza.  This deformation might be called planar splay or single splay.  For example, consider the director field
\begin{equation}
\hat{\bm{n}}(x,y,z)=\frac{(x,y,0)}{\sqrt{x^2+y^2}},
\end{equation}
which is shown in Fig.~2(a).  For this director field, explicit calculations give
\begin{align}
&\bm{B}=0,\quad
T=0,\quad
S=\frac{1}{\sqrt{x^2+y^2}},\\
&\bm{\Delta}=\frac{1}{2(x^2+y^2)^{3/2}}
\begin{pmatrix}
y^2 & -xy & 0\\
-xy & x^2 & 0\\
0   &   0 & -x^2-y^2
\end{pmatrix}
.\nonumber
\end{align}
This result shows that single splay is a linear combination of $S$ and $\bm{\Delta}$.  In particular, the tensor $\bm{\Delta}$ contains the directional information that identifies which of the directions perpendicular to the local $\hat{\bm{n}}$ is the splayed direction, and which is the uniform direction.

From Eq.~(\ref{oseenfranknew}), the free energy density has the components
\begin{align}
&F_S = \frac{1}{2}(K_{11}-K_{24})S^2 = \frac{K_{11}-K_{24}}{2(x^2+y^2)}, \nonumber\\
&F_\Delta=K_{24}\Tr(\bm{\Delta}^2) = \frac{K_{24}}{2(x^2+y^2)},
\end{align}
and hence the total
\begin{equation}
F = \frac{K_{11}}{2(x^2+y^2)}.
\end{equation}
We see that $K_{24}$ drops out of the free energy for single splay, precisely because single splay is a combination of $S$ and $\bm{\Delta}$.  Thus, $K_{11}$ is the relevant elastic constant for single splay.

This free energy expression has an important consequence for the splay Frederiks transition.  The experimental geometry for this transition is shown in Fig.~2(b).  We can see that this geometry has single splay, not double splay, because the director field stays in a plane.  Hence, the field-induced distortion is a combination of $S$ and $\bm{\Delta}$, and the relevant elastic constant is $K_{11}$.  Hence, the critical field for this transition is determined by $K_{11}$, not by $(K_{11}-K_{24})$.  Of course, this result is consistent with how the splay Frederiks transition has been analyzed for many years.

\subsection{Cholesteric (single) twist}

The same argument for single and double splay also applies to single and double twist.  In Sec.~II, we argued that a deformation with nonzero $T$ but zero $\bm{B}$, $S$, and $\bm{\Delta}$ is 3D double twist.  For comparison, we might consider a deformation with single twist, as in a cholesteric phase.  For example, consider the director field
\begin{equation}
\hat{\bm{n}}(x,y,z)=(\cos qz,\sin qz,0),
\end{equation}
which is shown in Fig.~2(c).  For this director field, we calculate
\begin{align}
&\bm{B}=0,\quad
T=-q,\quad
S=0,\\
&\bm{\Delta}=\frac{q}{2}
\begin{pmatrix}
0        & 0       & -\sin qz\\
0        & 0       &  \cos qz\\
-\sin qz & \cos qz &        0
\end{pmatrix}
.\nonumber
\end{align}
This result shows that cholesteric twist is a linear combination of $T$ and $\bm{\Delta}$.  The tensor $\bm{\Delta}$ contains the directional information about which of the directions perpendicular to the local $\hat{\bm{n}}$ is the helical axis, and which is the uniform direction.

From Eq.~(\ref{oseenfranknew}), the free energy density has the components
\begin{align}
&F_T = \frac{1}{2}(K_{22}-K_{24})T^2 = \frac{1}{2}(K_{22}-K_{24})q^2, \nonumber\\
&F_\Delta=K_{24}\Tr(\bm{\Delta}^2) = \frac{1}{2}K_{24}q^2,
\end{align}
and hence the total
\begin{equation}
F = \frac{1}{2}K_{22}q^2.
\end{equation}
Thus, $K_{24}$ drops out of the free energy for cholesteric twist, because it is a combination of $T$ and $\bm{\Delta}$, and $K_{22}$ is the relevant elastic constant for cholesteric twist.

The argument for the splay Frederiks transition also applies to the twist Frederiks transition.  As shown in Fig.~2(d), this transition involves single twist, not double twist.  Hence, the critical field is determined by $K_{22}$, not by $(K_{22}-K_{24})$.

\subsection{Cholesteric vs.\ blue phase}

The distinction between single and double twist provides a simple way to compare the free energies of cholesteric and blue phases in chiral liquid crystals.

If a liquid crystal is chiral, the free energy has the achiral nematic terms of Eq.~(\ref{oseenfranknew}), plus an additional chiral term proportional to the twist $T$.  The combination of the achiral and chiral terms favor  s a certain optimal value of $T$.  The question is:  How does the chiral liquid crystal achieve this twist?  Does it form a cholesteric or a blue phase?

In a cholesteric phase, the director field forms a helix with single twist everywhere.  This single twist is a combination of $T$ and $\bm{\Delta}$.  The mode $T$ provides a favorable, negative contribution to the free energy, but the mode $\bm{\Delta}$ provides an unfavorable, positive contribution to the free energy.

\begin{figure*}
\begin{tabular}{ccc}
(a) $\bm{\nabla}\cdot\hat{\bm{c}}=0$ & (b) $\bm{\nabla}\cdot\hat{\bm{c}}<0$ & (c) $\bm{\nabla}\cdot\hat{\bm{c}}>0$ \\
\includegraphics[width=5.8cm]{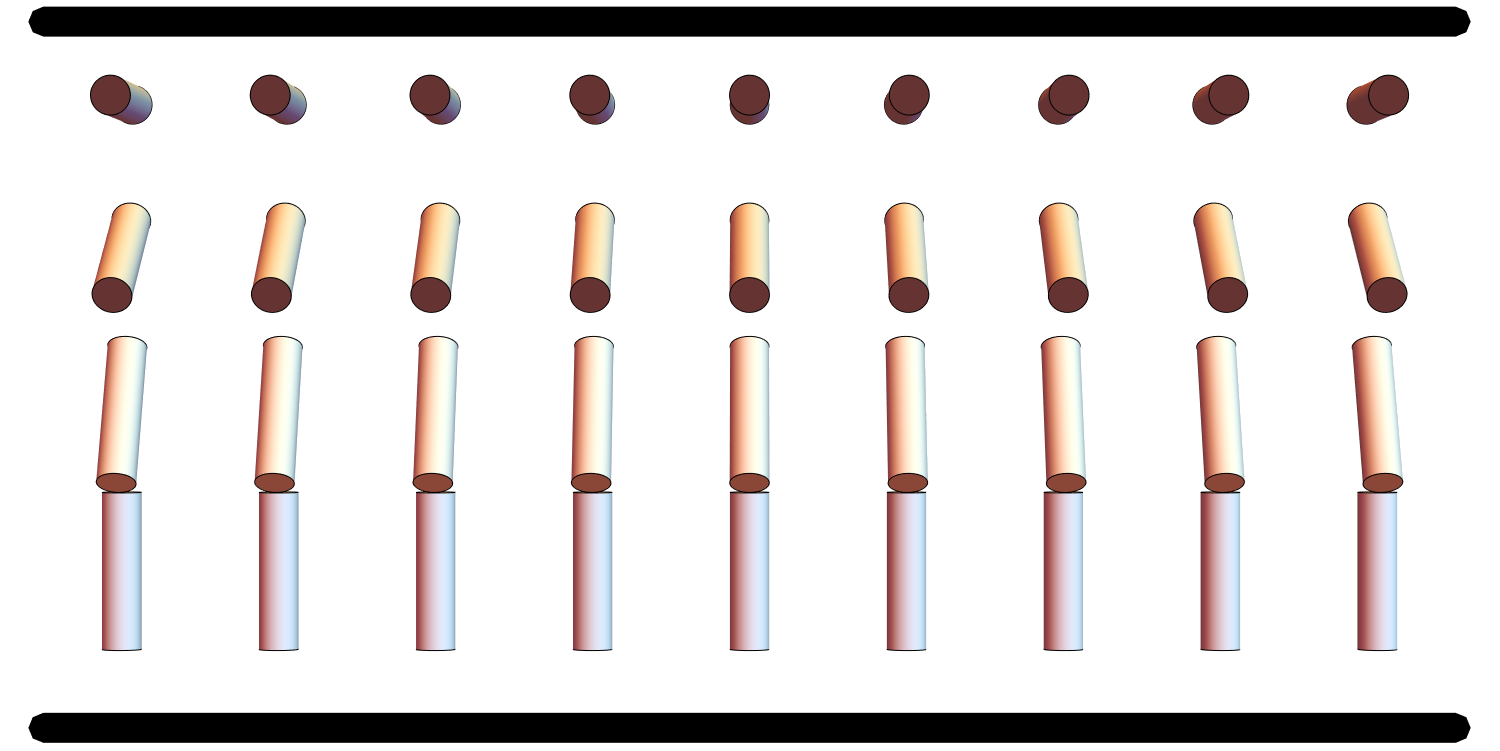} & \includegraphics[width=5.8cm]{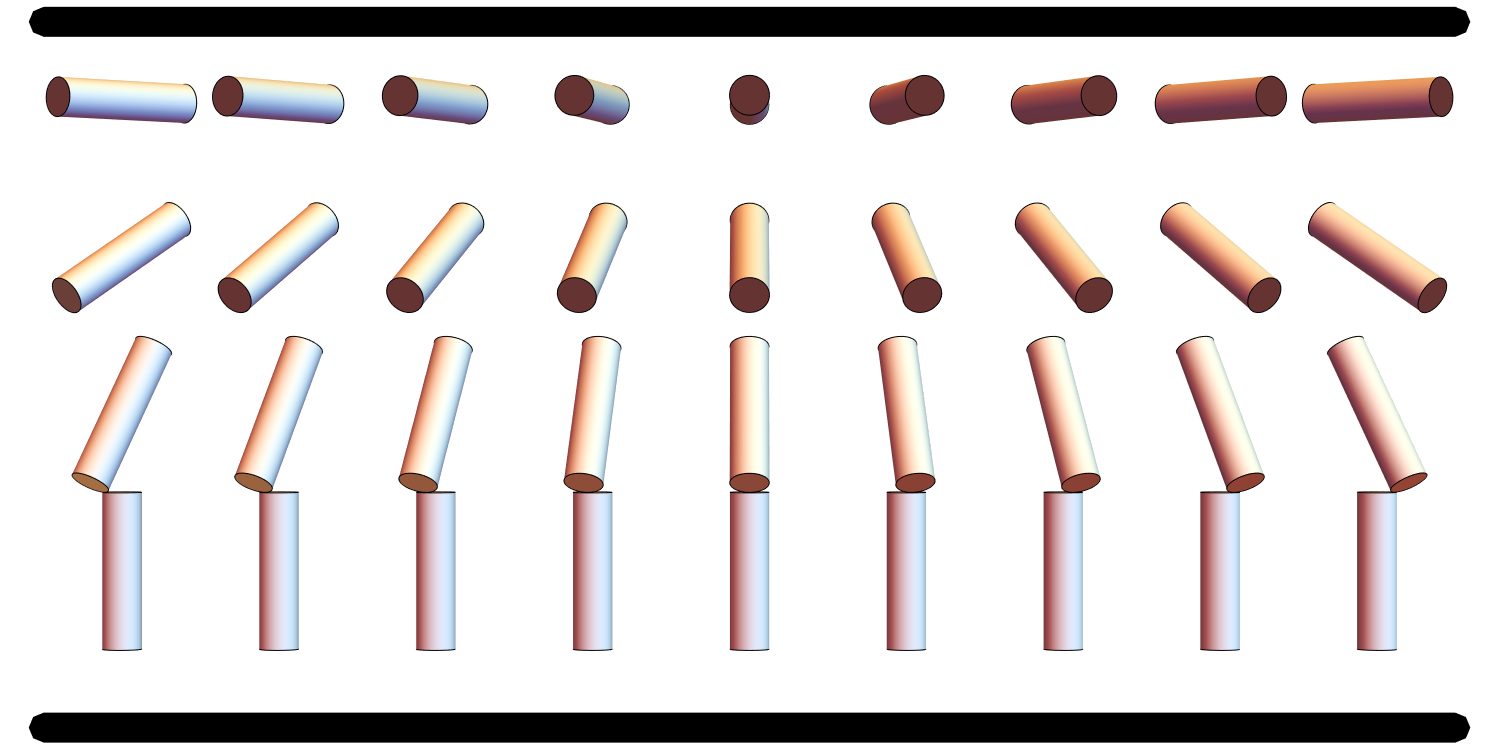} & \includegraphics[width=5.8cm]{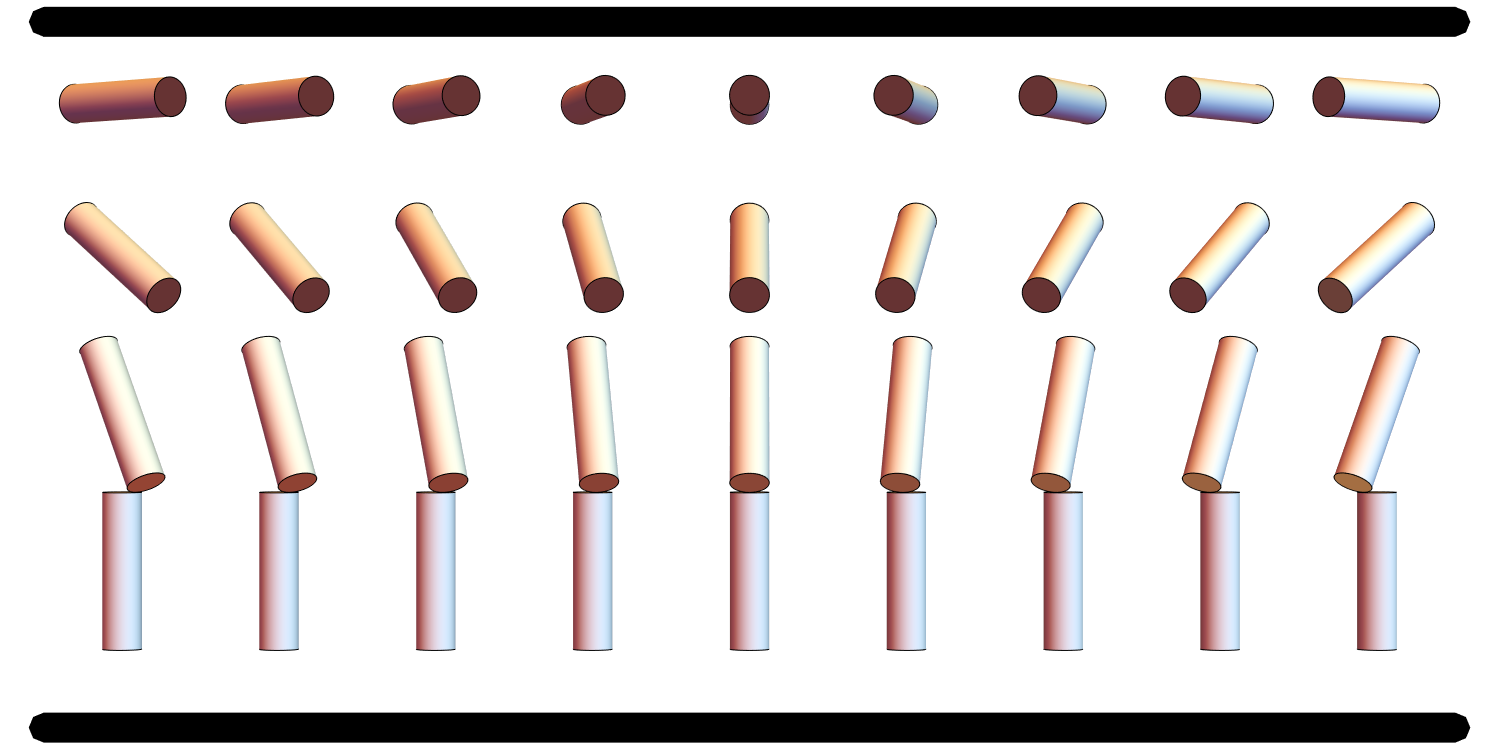} \\
\end{tabular}
\caption{(Color online) Director configurations in a hybrid aligned nematic liquid crystal.  (a)~Uniform top surface orientation $\hat{\bm{c}}(x,y)$, so that $\bm{\nabla}\cdot\hat{\bm{c}}=0$.  (b)~Modulation with $\bm{\nabla}\cdot\hat{\bm{c}}<0$.  This state has increased $S$ and reduced $\bm{\Delta}$, compared with the uniform state.  The increased $S$ can be seen clearly in the top middle of the director field.  (b)~Modulation with $\bm{\nabla}\cdot\hat{\bm{c}}>0$.  This state has increased $\bm{\Delta}$ and reduced $S$, compared with the uniform state.  The increased $\bm{\Delta}$ can be seen clearly in the top middle.}
\end{figure*}

In a blue phase, the director field forms a complex network of double twist tubes separated by disclination lines.  Because the tubes have double rather than single twist, they have pure $T$ with no $\bm{\Delta}$.  Hence, $T$ provides a favorable contribution to the free energy, while $\bm{\Delta}$ makes no contribution.  However, it is impossible to fill Euclidean space with double twist, as discussed below in Sec.~VI(B).  Rather, geometric considerations require a finite density of disclination lines.  These disclinations make an unfavorable contribution to the free energy.

Comparing these two possiblities, we can see that the favorable contributions to the free energy are the same, but the unfavorable contributions are different.  Thus, the relative stabilities of cholesteric and blue phases depends on which is worse:  the $\bm{\Delta}$ mode in a cholesteric phase, or the disclination lines in a blue phase.  That issue depends on the magnitude of $K_{24}$ compared with the free energy of the disclination cores, in which the liquid-crystal order is disrupted.  If the disclination core energy is high compared with $K_{24}$, then the liquid crystal will form a cholesteric phase.  If $K_{24}$ is high compared with the disclination core energy, then the liquid crystal will form a blue phase.

This conclusion that large $K_{24}$ is necessary to stabilize a blue phase is certainly not new.  It goes back to early work by Meiboom \emph{et al.}~\cite{Meiboom1981}, who considered saddle-splay as a surface free energy along the disclination lines as internal surfaces.  Related arguments have been made in the context of smectic blue phases~\cite{DiDonna2002,DiDonna2003} and double-tilt blue phases~\cite{Chakrabarti2006}.  We only suggest that this theoretical approach with modes $T$ and $\bm{\Delta}$ provides a particularly simple way to understand why $K_{24}$ is important.

\section{Further examples}

\subsection{Hybrid aligned nematic liquid crystal}

In the hybrid aligned nematic geometry, a liquid crystal is confined between two isotropic media, so that the director field has homeotropic (perpendicular) anchoring on the bottom and degenerate planar (tangential) anchoring on the top, or vice versa.  This type of system has been studied experimentally and theoretically over many years, as in Refs.~\cite{Sparavigna1994,Lavrentovich1995}.  Experimentally, these systems exhibit complex modulated structures in the director field.  Theoretically, the modulated structures have been explained by effects of saddle-splay, regarded as surface elasticity.  Here, we re-analyze the same system in terms of the four bulk modes discussed in this article.

Suppose that a cell extends from $z=0$ to $d$.  At the bottom surface $z=0$, there is homeotropic anchoring, so that $\hat{\bm{n}}(x,y,0)=\hat{\bm{z}}$.  At the top surface $z=d$, there is degenerate planar anchoring, so that $\hat{\bm{n}}$ must be in the $xy$ plane, and all orientations in the $xy$ plane have the same energy.  Hence, we can write $\hat{\bm{n}}(x,y,d)=\hat{\bm{c}}(x,y)$, where $\hat{\bm{c}}(x,y)$ is a unit vector in the $xy$ plane.  For the liquid crystal in the interior, an approximate form for the director field is
\begin{equation}
\hat{\bm{n}}(x,y,z)=\hat{\bm{c}}(x,y)\sin\frac{\pi z}{2d}+\hat{\bm{z}}\cos\frac{\pi z}{2d}.
\label{ninterior}
\end{equation}
This approximation is shown schematically in Fig.~3.  It is reasonable if $\hat{\bm{c}}(x,y)$ is slowly varying and the elastic constants are approximately equal.  The question is:  In the lowest-free-energy state, does the liquid crystal have a uniform surface orientation $\hat{\bm{c}}(x,y)$?  Or can the liquid crystal reduce its free energy with some modulation in $\hat{\bm{c}}(x,y)$?

\begin{figure*}
\begin{tabular}{ccc}
(a) $T=0$ & (b) $T<0$ & (c) $T>0$ \\
\includegraphics[width=5cm]{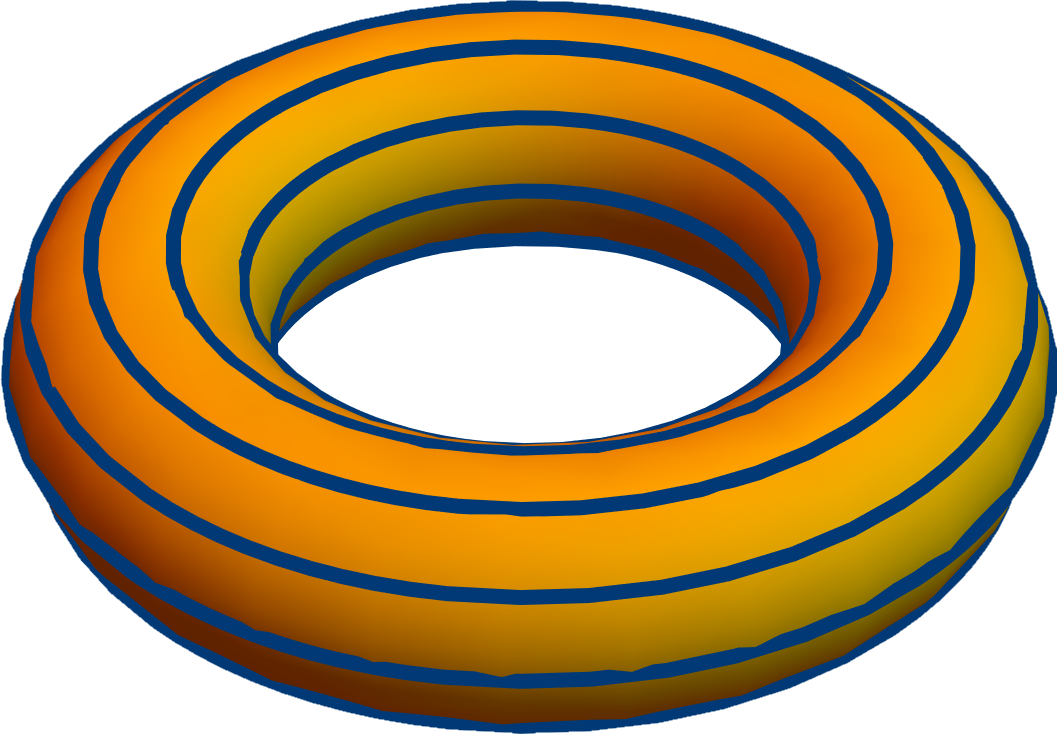} & \includegraphics[width=5cm]{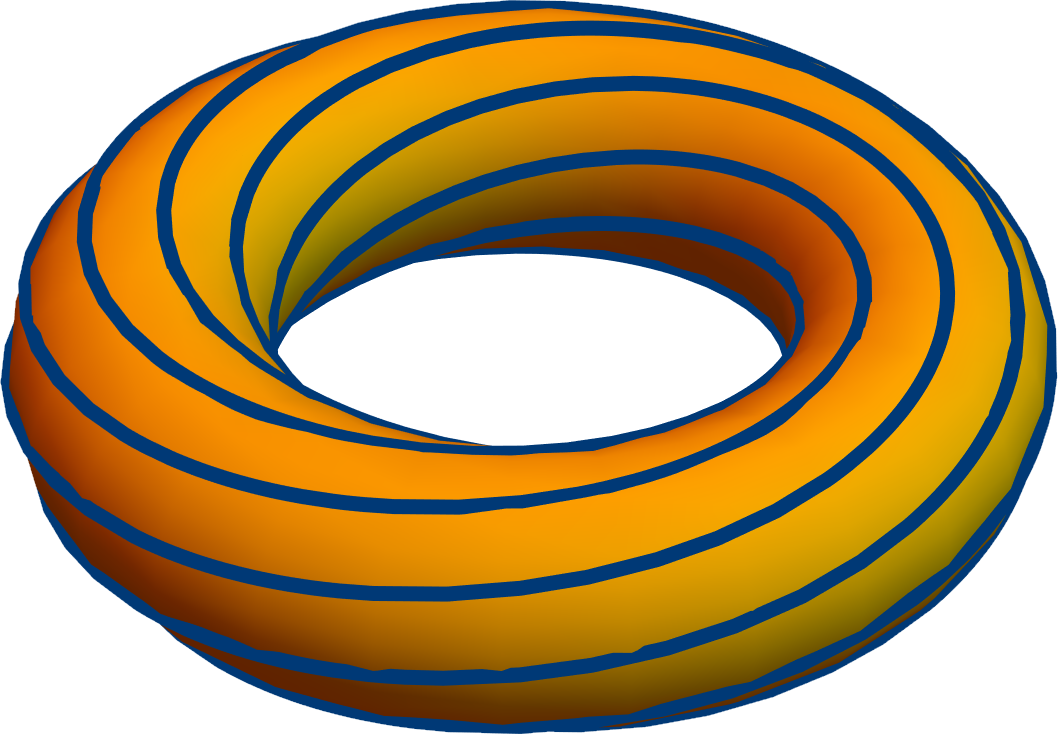} & \includegraphics[width=5cm]{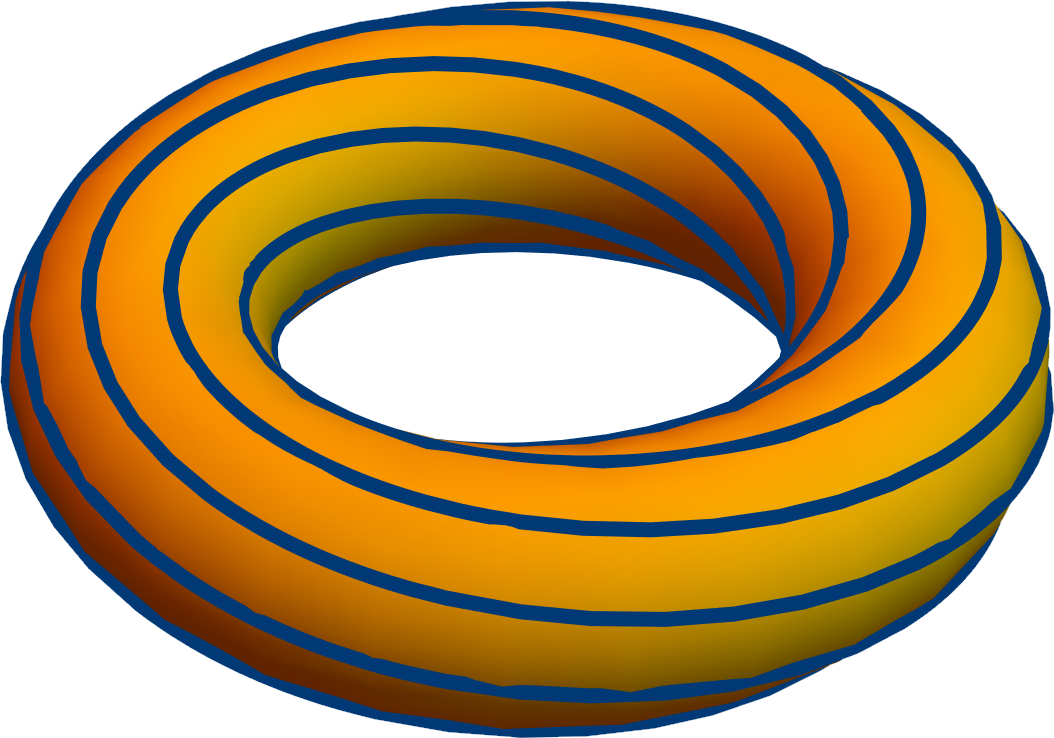} \\
\end{tabular}
\caption{(Color online) Director configurations for a nematic liquid crystal in a torus.  (a)~Achiral with $T=0$.  (b)~Chiral with $T<0$.  (c)~Chiral with $T>0$.  Figure is based on Ref.~\cite{Koning2014}.}
\end{figure*}

To answer that question, we use Eq.~(\ref{ninterior}) for the 3D director field to calculate the 3D distortions $S$, $T$, $\bm{B}$, and $\bm{\Delta}$.  We put those distortions into Eq.~(\ref{oseenfranknew}) to calculate the 3D Frank free energy density.  We then integrate over $z=0$ to $d$ to calculate the effective 2D Frank free energy density, $F_{2D}(x,y)=\int_0^d dz F(x,y,z)$, in terms of the surface orientation $\hat{\bm{c}}(x,y)$.
This calculation gives the four components
\begingroup
\allowdisplaybreaks
\begin{align}
&F_{2D}^B=K_{33}\left[\frac{\pi^2}{16d}
+\frac{3d}{16}(\bm\nabla\times\hat{\bm{c}})^2\right],\\
&F_{2D}^T=(K_{22} -K_{24})\left[\frac{d}{16}(\bm\nabla\times\hat{\bm{c}})^2\right],\nonumber\\
&F_{2D}^S=(K_{11} -K_{24})\left[\frac{\pi^2}{16d}
-\frac{\pi}{4}\bm\nabla\cdot\hat{\bm{c}}
+\frac{d}{4}(\bm\nabla\cdot\hat{\bm{c}})^2\right],\nonumber\\
&F_{2D}^\Delta=K_{24}\left[\frac{\pi^2}{16d}
+\frac{\pi}{4}\bm\nabla\cdot\hat{\bm{c}}
+\frac{d}{4}(\bm\nabla\cdot\hat{\bm{c}})^2
+\frac{d}{16}(\bm\nabla\times\hat{\bm{c}})^2\right],\nonumber
\end{align}
where $\bm\nabla\cdot\hat{\bm{c}}$ and $\bm\nabla\times\hat{\bm{c}}$ are the 2D divergence and curl of the surface orientation, respectively.  As a result, the total effective 2D free energy density becomes
\begin{align}
F_{2D}=&\frac{\pi^2 (K_{11} +K_{33})}{16d}
-\frac{\pi(K_{11} -2K_{24})}{4}\bm\nabla\cdot\hat{\bm{c}}\\
&+\frac{K_{11} d}{4}(\bm\nabla\cdot\hat{\bm{c}})^2
+\frac{(K_{22} +3K_{33})d}{16}(\bm\nabla\times\hat{\bm{c}})^2.\nonumber
\end{align}
\endgroup

To interpret these expressions, the most important feature to notice is the linear dependence on $\bm\nabla\cdot\hat{\bm{c}}$.  First consider a state with uniform surface orientation, so that $\bm\nabla\cdot\hat{\bm{c}}=0$, as in Fig.~3(a).  This state is loaded with three of the modes, $S$, $\bm{B}$, and $\bm{\Delta}$, which all contribute to the free energy.  Now suppose the surface orientation has a small variation with $\bm\nabla\cdot\hat{\bm{c}}<0$, as in Fig.~3(b).  This variation increases the $S$ and reduces the $\bm{\Delta}$, compared with the uniform state.  Hence, this variation reduces the free energy compared with the uniform state, provided that the splay coefficient $(K_{11}-K_{24})$ is less than the biaxial splay coefficient $K_{24}$.  By comparison, suppose the surface orientation has a small variation with $\bm\nabla\cdot\hat{\bm{c}}>0$, as in Fig.~3(c).  This variation reduces the $S$ and increases the $\Delta$, compared with the uniform state.  Hence, that variation reduces the free energy compared with the uniform state, provided that $(K_{11}-K_{24})>K_{24}$.

From this argument, we see that the lowest-free-energy state has uniform surface orientation only in the special case that $(K_{11}-K_{24})=K_{24}$, or $K_{11}=2K_{24}$.  Interestingly, that is exactly the case of equal elastic constants, discussed in Sec.~III.  Although that case is a theoretical possibility, an experimental liquid crystal will normally have elastic constants that are at least slightly different.  If $K_{11}<2K_{24}$, then the energetic benefit of reducing $\Delta$ exceeds the energetic cost of increasing $S$, and hence the system can reduce its free energy by going to a modulation with $\bm\nabla\cdot\hat{\bm{c}}<0$.  Conversely, if $K_{11}>2K_{24}$, then the energetic benefit of reducing $S$ exceeds the energetic cost of increasing $\Delta$, and hence the system can reduce its free energy by going to a modulation with $\bm\nabla\cdot\hat{\bm{c}}>0$.

This substitution of $S$ for $\Delta$, or vice versa, can be regarded as the origin for the complex textures that are observed in hybrid aligned nematic liquid crystals.  Of course, there is no contradiction between this interpretation and previous theories based on surface elasticity; they are just two ways of describing the same behavior.

\subsection{Liquid crystal in torus}

Suppose we have a nematic liquid crystal inside a torus, with degenerate planar anchoring on the surface.  Does the director field form a simple, achiral configuration, running along the long axis of the torus, as shown in Fig.~4(a)?  Or does it break reflection symmetry and form a chiral configuration, with a double twist from the central axis to the surface of the torus, as shown in Fig.~4(b,c)?

This system was investigated theoretically in Ref.~\cite{Koning2014}, and further, more mathematically, in Ref.~\cite{Pedrini2018}.  Those studies use the perspective of saddle-splay as surface elasticity, and show that it favors alignment of the director field along the highly curved direction on the surface of the torus.  If $K_{24}$ is small, then the bulk free energy favoring alignment along the long axis exceeds the surface free energy favoring alignment in the highly curved direction, and the director field forms the simple, achiral configuration.  However, if $K_{24}$ is large enough, then the surface free energy exceeds the bulk free energy, and the director field forms a chiral configuration.

\begin{figure*}
\begin{tabular}{ccc}
(a) $T=0$ & (b) $T<0$ & (c) $T>0$ \\
\includegraphics[width=5cm]{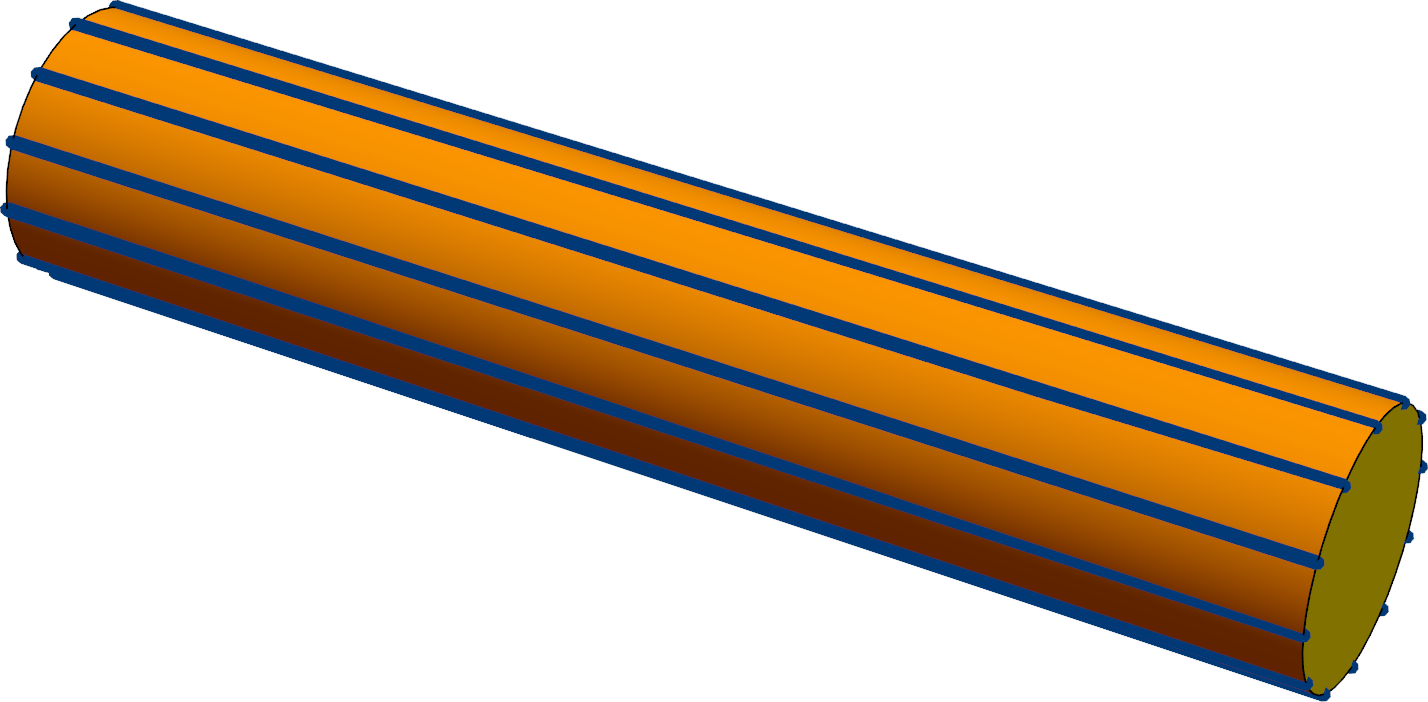} & \includegraphics[width=5cm]{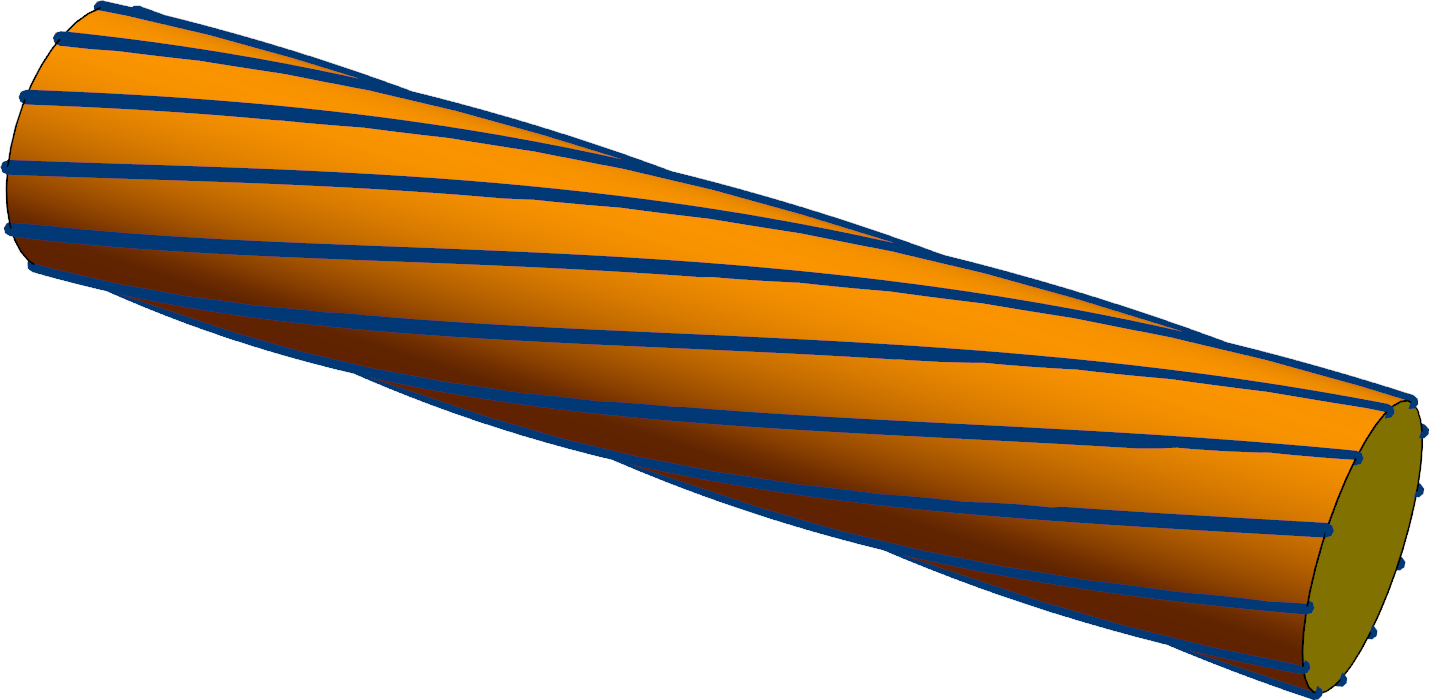} & \includegraphics[width=5cm]{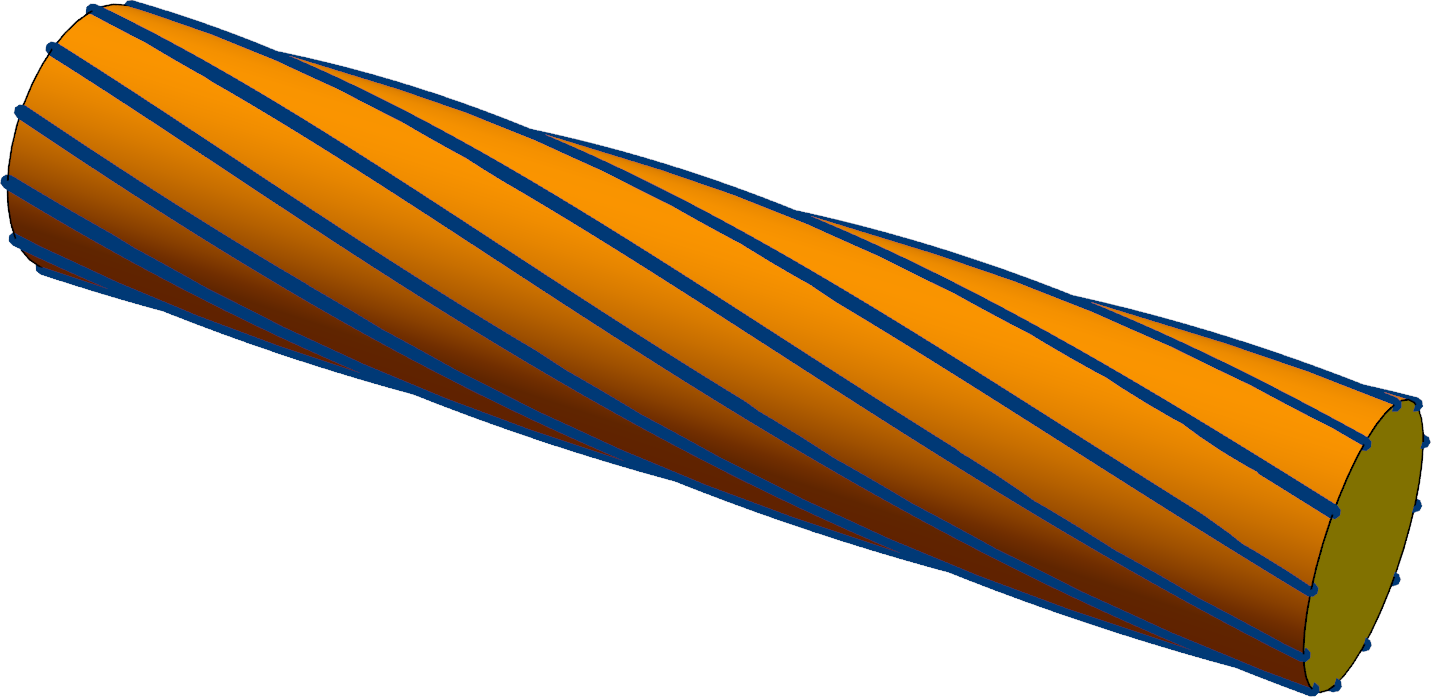} \\
\end{tabular}
\caption{(Color online) Director configurations for a nematic liquid crystal in a cylinder.  (a)~Achiral with $T=0$.  (b)~Chiral with $T<0$.  (c)~Chiral with $T>0$.}
\end{figure*}

Here, we consider the same problem in terms of the four bulk deformation modes.  For this calculation, we use the same toroidal coordinate system and the same director ansatz as Ref.~\cite{Koning2014}.  We assume that their chiral order parameter $\omega$ is small, and expand the free energy in powers of $\omega$.  The four terms in the free energy, integrated over the interior of the torus, then become
\begingroup
\allowdisplaybreaks
\begin{align}
& F_B = K_{33}\pi^2 R_1\biggl[2\left(1-\frac{(\xi^2 -1)^{1/2}}{\xi}\right)\nonumber\\
&\qquad-\left(6\xi-\frac{6\xi^4 -9\xi^2 +1}{(\xi^2 -1)^{3/2}}\right)\xi\omega^2 +O(\omega^4)\biggr],\nonumber\\
& F_T = (K_{22}-K_{24})\pi^2 R_1 \frac{4\xi^3}{(\xi^2 -1)^{3/2}}\omega^2 +O(\omega^6),\nonumber\\
& F_S = 0, \quad F_\Delta = O(\omega^6),
\label{ftorus}
\end{align}
\endgroup
where $\xi=R_1/R_2>1$ is the aspect ratio of the torus.  From these expressions, we see that the splay free energy is exactly zero (by construction), and the $\bm{\Delta}$ mode free energy is approximately zero, so the important physics arises from the competition between bend and twist.  In the achiral configuration with $\omega=0$, the bend free energy is high, and the twist free energy is zero.  As the magnitude of $\omega$ increases, the bend free energy decreases and the twist free energy increases.  This trade-off might or might not reduce the total free energy, depending on the total coefficient of $\omega^2$.  From Eq.~(\ref{ftorus}), we see that this coefficient is negative, and hence a chiral configuration is favored, provided that
\begin{align}
\frac{K_{22}-K_{24}}{K_{33}}&<\frac{6\xi(\xi^2 -1)^{3/2} -6\xi^4 +9\xi^2 -1}{4\xi^2}\nonumber\\
&\approx\frac{5}{16\xi^2}\text{ if }\xi\gg1.
\label{torusinequality}
\end{align}
This result is exactly the same critical threshold found in Ref.~\cite{Koning2014}.  Thus, we see that their result can be understood from the competition between the bulk free energies of bend and twist, without reference to any surface alignment.  We note that the $\bm{\Delta}$ mode is not involved with this competition.  Rather, $K_{24}$ enters into the prediction because the twist is double twist, and the elastic constant for double twist is $(K_{22}-K_{24})$.

\subsection{Chromonic liquid crystal in cylinder}

Let us apply the results from the torus to the simpler case of a liquid crystal in a cylinder.  We can ask whether the director field forms a simple uniform configuration parallel to the cylinder axis, as in Fig.~5(a), or whether it breaks reflection symmetry and forms a chiral configuration, as in Fig.~5(b,c).

A cylinder can be regarded as a limiting case of a torus, in the limit where $R_1\to\infty$ and $R_2$ remains finite, so that $\xi\to\infty$.  In that case, Eq.~(\ref{torusinequality}) implies that a chiral configuration is favored if $(K_{22}-K_{24})/K_{33}<0$.  Hence, we must ask:  Is it possible for a liquid crystal to have $K_{22}<K_{24}$?

This issue was addressed by Ericksen~\cite{Ericksen1966}, who developed several inequalities for liquid-crystal elasticity.  Translated into our current notation, Ericksen's argument is essentially:  We assume that a uniform nematic state is stable, so any gradients of the director must cost free energy, and hence all of the elastic coefficients in Eq.~(\ref{oseenfranknew}) must be positive.  Thus, he concludes that $0<K_{24}<K_{11}$, $0<K_{24}<K_{22}$, and $0<K_{33}$.  If these inequalities are correct, then the cylinder never has a transition from achiral to chiral; the uniform, achiral state is always stable.  However, this reasoning has a flaw:  We are concluding that the uniform state is stable based on a theory that \emph{assumes} the uniform state is stable.  We cannot draw any conclusions from this circular argument.

Surprisingly, there are certain liquid crystal materials that violate the Ericksen inequalities, and actually exhibit $K_{22}<K_{24}$.  In particular, several studies have investigated the lyotropic chromonic liquid crystals Sunset Yellow (SSY) and disodium cromoglycate (DSCG) in cylindrical capillaries~\cite{Davidson2015,Nayani2015,Fu2017} or cylindrical shells~\cite{Javadi2018}.  Experimentally, those studies show that the uniform, achiral state can become unstable to the formation of a chiral state.  Theoretically, they model the transition in terms of saddle-splay as surface elasticity, which favors alignment of the director field along the curved direction on the surface of the torus.

Here, we see that this symmetry-breaking transition has a simple interpretation in terms of the double twist $T$.  When $K_{22}<K_{24}$, the coefficient of $T^2$ in the free energy density~(\ref{oseenfranknew}) becomes negative.  Hence, the state with $T=0$ becomes unstable to the formation of $T\not=0$, which can be either positive or negative.  The director field will then form a configuration determined by the combination of the favorable free energy associated with double twist and the unfavorable free energy associated with bend.

Based on this argument, there is an analogy between the double twist instability of chromonic liquid crystals and the bend instability of bent-core liquid crystals.  Dozov~\cite{Dozov2001} argued that the bend elastic constant $K_{33}$ of bent-core liquid crystals can become negative, leading to the formation of a nonuniform phase, and such phases have been found experimentally; see the discussion in Ref.~\cite{Jakli2018}.  In chromonic liquid crystals, the double twist elastic constant $(K_{22}-K_{24})$ can become negative, which also leads to the formation of a nonuniform phase.  In both cases, the nonuniform phase has a combination of bend and double twist, and the double twist is randomly right- or left-handed.

If the coefficient of $T^2$ in the free energy density becomes negative, then we need some mechanism to stabilize the free energy so that it cannot decrease without limit.  The free energy might be stabilized by a higher-order power, such as $T^4$, or a higher derivative of the director field $\hat{\bm{n}}$, as discussed below in Sec.~VI(A).  Alternatively, the free energy might be stabilized by a compatibility constraint, as discussed in Sec.~VI(B).   

\subsection{Curvature of smectic layers}

In the smectic-A liquid crystal phase, the molecules lie in (approximately) equally spaced layers, and the director field is normal to the layers.  The layers are curved surfaces, which can be treated through the methods of differential geometry.  Hence, the deformations of the director field can be related to the curvature of the layers.  In particular, the splay of the director field is twice the mean curvature of the layers, and the saddle-splay is twice the Gaussian curvature~\cite{Kleman2003}.

Here, because we are analyzing director deformations in terms of the four modes $\bm{B}$, $T$, $S$, and $\bm{\Delta}$, we should determine how all four of these modes are related to the layer curvature.

Suppose the equilibrium smectic layers are in the $xy$ plane, and there are small local displacements from $z$ to $z+u(x,y,z)$.  To lowest nontrivial order in $u$, the director field is
\begin{equation}
\hat{\bm{n}}=\pm(\hat{\bm{z}}-\bm{\nabla}_\perp u).
\end{equation}
Hence, the director deformations become
\begin{align}
&\bm{B}=\bm{\nabla}_\perp (\partial_z u),\quad
T=0,\quad
S=\mp\nabla_\perp^2 u,\\
&\bm{\Delta}=\pm\begin{pmatrix}
\frac{1}{2}(\partial_y^2 u -\partial_x^2 u)  & -\partial_x \partial_y u & 0\\
-\partial_x \partial_y u & \frac{1}{2}(\partial_x^2 u -\partial_y^2 u) & 0\\
0 & 0 & 0
\end{pmatrix},\nonumber
\end{align}
with the $\pm$ signs depending on the arbitrary choice of $\hat{\bm{n}}$ or $-\hat{\bm{n}}$.  The twist is zero, as it must be for a director field that is perpendicular to layers.  The bend is the perpendicular gradient of $\partial_z u$, which is the variation in layer spacing.  If the layers are equally spaced, then the bend is also zero.  Hence, the important two deformations are the splay and biaxial splay.

We can relate these properties to the curvature of the smectic layers.  To lowest nontrivial order in $u$, the curvature tensor is
\begin{equation}
K_\alpha^\beta =
\begin{pmatrix}
-\partial_x^2 u & -\partial_x \partial_y u \\
-\partial_x \partial_y u & -\partial_y^2 u \\
\end{pmatrix}.
\end{equation}
If $\kappa_1$ and $\kappa_2$ are the two principal curvatures, then the mean curvature becomes
\begin{equation}
\frac{1}{2}(\kappa_1 +\kappa_2)=\frac{1}{2}K_\alpha^\alpha = -\frac{1}{2}\nabla_\perp^2 u = \pm\frac{1}{2}S.
\end{equation}
Hence, the geometric meaning of the splay $S$ is twice the mean curvature.  Likewise, the traceless part of the curvature tensor becomes
\begin{equation}
K_\alpha^\beta - \frac{1}{2}K_\gamma^\gamma \delta_\alpha^\beta =
\begin{pmatrix}
\frac{1}{2}(\partial_y^2 u -\partial_x^2 u)  & -\partial_x \partial_y u \\
-\partial_x \partial_y u & \frac{1}{2}(\partial_x^2 u -\partial_y^2 u)
\end{pmatrix}.
\end{equation}
Hence, the geometric meaning of the biaxial splay tensor $\bm{\Delta}$ is the traceless part of the curvature tensor.  Its eigenvalues are $\pm\frac{1}{2}(\kappa_1 -\kappa_2)$, and its eigenvectors are the principal curvature directions, along with a third eigenvalue of zero corresponding the eigenvector $\hat{\bm{n}}$.

From Eq.~(\ref{saddlesplay}), the standard saddle-splay becomes
\begin{align}
&\bm{\nabla}\cdot\left[\hat{\bm{n}}(\bm{\nabla}\cdot\hat{\bm{n}})+\hat{\bm{n}}\times(\bm{\nabla}\times\hat{\bm{n}})\right]
=\frac{1}{2}S^2 +\frac{1}{2}T^2 - \Tr(\bm{\Delta}^2)\nonumber\\
&\quad=\frac{1}{2}(\kappa_1 +\kappa_2)^2 - \frac{1}{2}(\kappa_1 -\kappa_2)^2 = 2\kappa_1 \kappa_2.
\end{align}
This agrees with the textbook result that the saddle-splay is twice the Gaussian curvature.

We have verified that these results hold exactly, beyond the approximation of small layer displacements, for a toroidal focal conic structure.

\section{Related issues}

\subsection{Second-derivative elasticity and $K_{13}$}

In the literature on elasticity of liquid crystals, the Oseen-Frank free energy density is sometimes written with two divergence terms~\cite{Nehring1971,Nehring1972,Kleman2003},
\begin{align}
F=&\frac{1}{2}K_{11}S^2 + \frac{1}{2}K_{22}T^2 + \frac{1}{2}K_{33}|\bm{B}|^2 \\
&-K_{24}\bm{\nabla}\cdot\left[\hat{\bm{n}}(\bm{\nabla}\cdot\hat{\bm{n}})+\hat{\bm{n}}\times(\bm{\nabla}\times\hat{\bm{n}})\right]\nonumber\\
&+K_{13}\bm{\nabla}\cdot\left[\hat{\bm{n}}(\bm{\nabla}\cdot\hat{\bm{n}})\right].\nonumber
\end{align}
The origin and effects of the $K_{13}$ term have been analyzed in detail~\cite{Pergamenshchik1998,Pergamenshchik1999,Pergamenshchik2000}.  In this article, we have discussed how the $K_{24}$ term can be written in terms of the modes $S$, $T$, and $\bm{\Delta}$.  Can the $K_{13}$ term be written in a similar way?

To answer that question, we note that the $K_{24}$ and $K_{13}$ terms are actually different types of mathematical objects.  Although the $K_{24}$ saddle-splay term appears superficially as if it includes second derivatives of $\hat{\bm{n}}$, the second derivatives of the first piece exactly cancel the second derivatives of the second piece.  As a result, this term depends only on first derivatives of $\hat{\bm{n}}$.  That is why it can be expressed in terms of $S$, $T$, and $\bm{\Delta}$, which are all combinations of first derivatives of $\hat{\bm{n}}$.  Hence, it is quite appropriate to include this term in the Oseen-Frank free energy, along with the $K_{11}$, $K_{22}$, and $K_{33}$ terms, which also involve first derivatives of $\hat{\bm{n}}$.

By contrast, in the $K_{13}$ term, the second derivatives of $\hat{\bm{n}}$ do not cancel.  Instead, that term becomes
\begin{align}
\bm{\nabla}\cdot\left[\hat{\bm{n}}(\bm{\nabla}\cdot\hat{\bm{n}})\right]
&=\bm{\nabla}\cdot\left[\hat{\bm{n}}S\right]
=S^2 + (\hat{\bm{n}}\cdot\bm{\nabla})S\nonumber\\
&=(\partial_i n_i)(\partial_j n_j) + n_i \partial_i \partial_j n_j.
\end{align}
If we include this term in the free energy, with no other second-derivative terms, then the theory would become unstable, because it would favor arbitrary large second derivatives of $\hat{\bm{n}}$.  To avoid that problem, we would need to add other second-derivative terms, such as $(\partial_i \partial_j n_k)(\partial_i \partial_j n_k)$, to stabilize the free energy.  We recognize that such terms are formally smaller than other terms in the free energy, because they include more derivatives, but still they are necessary for stability.

From this discussion, we can see two reasonable options.  First, we might consider only first derivatives of $\hat{\bm{n}}$ in the theory.  In that case, the free energy would include the $K_{11}$, $K_{22}$, $K_{33}$, and $K_{24}$ terms, but not $K_{13}$.  Alternatively, we might develop a higher-order elasticity theory that includes all second derivatives of $\hat{\bm{n}}$.  For this higher-order elasticity, we could begin by decomposing the tensor $\partial_i \partial_j n_k$ into its normal modes (by analogy with the calculation in Sec.~II) and expressing the general second-derivative free energy in terms of those modes (by analogy with Sec.~III).  That analysis is beyond the scope of this article.

For almost all liquid crystal physics problems, the first option is sufficient.  However, there may be a few unusual problems where second-derivative elasticity is needed.  One example might be the chromonic liquid crystals discussed in Sec.~V(C), where the double twist elastic constant $(K_{22}-K_{24})$ becomes negative.  Another example might be the bend instability of bent-core liquid crystals, where the effective bend elastic constant $K_{33}$ becomes negative~\cite{Dozov2001,Jakli2018}.  In those unusual cases, second-derivative elasticity might be important for stabilizing the free energy.

\subsection{Compatibility}

In general, some director deformations can exist everywhere in space, and other deformations cannot.  For example, it is possible to fill up space with cholesteric single twist (which is a specific combination of $T$ and $\bm{\Delta}$), but it is impossible to fill up space with double twist (pure $T$).  That is the reason why blue phases must have tubes of double twist separated by disclination lines, rather than uniform double twist~\cite{Sethna1983}.  Similarly, it is possible to fill up space with pure bend of varying magnitude, as in Sec.~II(B1), or pure splay of varying magnitude, as in the hedgehog of Sec.~II(B3), but it is impossible to fill up space with pure bend or splay of constant magnitude.

This issue can be regarded as a \emph{compatibility} problem.  For any director field $\hat{\bm{n}}$, we can calculate derivatives to define the modes $\bm{B}$, $T$, $S$, and $\bm{\Delta}$.  However, that procedure does not work in reverse:  We cannot begin with any arbitrary set of $\bm{B}$, $T$, $S$, and $\bm{\Delta}$, and calculate the corresponding director field.  Only certain combinations of $\bm{B}$, $T$, $S$, and $\bm{\Delta}$ can be constructed from the same $\hat{\bm{n}}$.  Those combinations can be called compatible, while other combinations are incompatible.

An analogous issue of compatibility occurs in the theory of elastic solids.  For any displacement field $\bm{u}$, we can calculate derivatives to define the strain tensor $\bm{\epsilon}$.  However, we cannot begin with any arbitrary strain tensor and calculate a corresponding displacement field.  In this sense, the displacement field of an elastic solid is analogous to the director field of a liquid crystal, and the strain tensor is analogous to the liquid crystal deformation modes $\bm{B}$, $T$, $S$, and $\bm{\Delta}$.

The issue of compatibility has been studied extensively in the theory of elastic solids.  It is known that the strain tensor must satisfy certain constraints in order to be compatible with a displacement field.  By contrast, this issue has not been studied much in the theory of liquid crystals.  To our knowledge, it has only been investigated theoretically by Niv and Efrati~\cite{Niv2018}, for the case of 2D liquid crystals.  They derived the compatibility constraint, which depends on the Gaussian curvature of the 2D surface in which the liquid crystal exists.  For a flat surface, it is possible to have a simple case of a uniform liquid crystal, with zero bend and zero splay everywhere, but it is impossible to have uniform nonzero bend or uniform nonzero splay.  For a positively curved surface, such as a sphere, it is impossible to have even uniform zero bend or uniform zero splay.  By contrast, for a negatively curved surface, such as a saddle, it is possible to have uniform nonzero bend or uniform nonzero splay.

So far, the corresponding compatibility constraint or constraints have not yet been derived for 3D liquid crystals.  We anticipate that this calculation will be done in the future, and then it will establish what combinations of $\bm{B}$, $T$, $S$, and $\bm{\Delta}$ are compatible with 3D Euclidean flat space.

A compatibility constraint may be important for understanding the stability of the Oseen-Frank free energy in the case where one of the elastic constants becomes negative.  The free energy does not need to be positive-definite for all variations of $\bm{B}$, $T$, $S$, and $\bm{\Delta}$.  Rather, it only needs to be positive-definite for all \emph{compatible} variations of $\bm{B}$, $T$, $S$, and $\bm{\Delta}$, so that it will be positive-definite for all possible variations of $\hat{\bm{n}}$.  A mathematical constraint should help to determine what are the compatible variations.

\section{Conclusions}

In this article, we have presented a theoretical formalism to analyze director deformations in liquid crystals.  This formalism is based on a mathematical construction by Machon and Alexander~\cite{Machon2016}, which decomposes the director gradient tensor $\partial_i n_j$ into four modes:  splay, twist, bend, and $\bm{\Delta}$.  We re-express the Oseen-Frank free energy in terms of these modes, and show that it takes a simple form as the sum of squares.  Using that expression for the free energy, we re-analyze several previous problems in liquid-crystal physics, and show how they can be understood based on the four modes.

The main difference between our current approach and previous work is that we now regard all four modes as bulk elastic modes, while previous work usually considered saddle-splay as surface elasticity.  We emphasize that there is no contradiction between these perspectives.  Indeed, the theories are mathematically equivalent, and give the same predictions for experiments.  However, we suggest that the current approach provides a simpler and more intuitive way to understand the role of saddle-splay, and hence provides a useful tool for future theoretical research.

\acknowledgments

We would like to thank E. Efrati, A. J\'{a}kli, O. D. Lavrentovich, R. Mosseri, and J.-F. Sadoc for helpful discussions.  This work was supported by National Science Foundation Grant DMR-1409658.  Part of this work was performed at the Aspen Center for Physics, which is supported by National Science Foundation Grant PHY-1607761.

\appendix

\section{Calculation of four modes from tensor~$\bm{Q}$}

One application of the formalism discussed in this article may be to analyze simulations of complex liquid-crystal structures, such as skyrmions, half-skyrmions, and blue phases.  Such simulations are often done using the nematic order tensor field $\bm{Q}(\bm{r})$, rather than the director field $\hat{\bm{n}}(\bm{r})$.  For use in these simulations, we would like to express the four modes $\bm{B}$, $T$, $S$, and $\bm{\Delta}$ in terms of the nematic order tensor.

The nematic order tensor field is usually written as the traceless form
\begin{equation}
Q_{ij}=s\left(\frac{3}{2}n_i n_j -\frac{1}{2}\delta_{ij}\right),
\end{equation}
where $s$ is the scalar order parameter, i.e.\ the magnitude of nematic order, not to be confused with the splay $S$.  In any defect-free region where $s$ is approximately constant, we can just work with the tensor
\begin{equation}
q_{ij}=n_i n_j.
\end{equation}
It is then straightforward to convert $Q_{ij}=s(\frac{3}{2}q_{ij}-\frac{1}{2}\delta_{ij})$, or $q_{ij}=\frac{1}{3}(\delta_{ij}+2Q_{ij}/s)$.

Because $\hat{\bm{n}}$ is a unit vector, we have
\begin{equation}
n_i \partial_j n_k = q_{il} \partial_j q_{kl}.
\end{equation}
By making appropriate contractions of that equation, we can derive the bend vector
\begin{equation}
B_k = -n_i \partial_i n_k = -q_{il} \partial_i q_{kl},
\end{equation}
and the twist pseudoscalar
\begin{equation}
T=\epsilon_{ijk} n_i \partial_j n_k =\epsilon_{ijk} q_{il} \partial_j q_{kl}.
\end{equation}
It is impossible to define the splay scalar $S=\bm\nabla\cdot\hat{\bm n}$ uniquely in terms of $q_{ij}$, because $S$ is odd in $\hat{\bm{n}}$; i.e., it depends on the arbitrary choice of $\hat{\bm{n}}$ or $-\hat{\bm{n}}$.  However, we can uniquely define the splay vector $\bm{S}=S\hat{\bm{n}}$, because it is even in $\hat{\bm{n}}$.  The splay vector becomes
\begin{equation}
S_i = S n_i = n_i \partial_j n_j = q_{il} \partial_j q_{jl}.
\end{equation}
Likewise, it is impossible to define the second-rank tensor $\Delta_{ij}$ uniquely in terms of $q_{ij}$, because $\Delta_{ij}$ is also odd in $\hat{\bm{n}}$.  Instead, we can define the third-rank tensor
\begin{align}
\Delta_{ij} n_k = & \frac{1}{2} [n_k \partial_i n_j + n_k \partial_j n_i + n_k n_i B_j + n_k n_j B_i \nonumber\\
&\quad - S n_k (\delta_{ij} - n_i n_j)] \nonumber\\
= & \frac{1}{2} [q_{kl} \partial_i q_{jl} + q_{kl} \partial_j q_{il} + q_{ki} B_j + q_{kj} B_i \nonumber\\
&\quad - S_k (\delta_{ij} - q_{ij})].
\end{align}

\section{Director gradient modes in 2D}

In Sec.~II(A), we decompose the director gradient tensor $\partial_i n_j$ into the four modes $\bm{B}$, $T$, $S$, and $\bm{\Delta}$ in 3D.  Some researchers also investigate nematic liquid crystals in 2D.  In this appendix, we do the analogous decomposition in 2D, for use in such studies.

First, consider the number of degrees of freedom.  In 2D, the tensor $\partial_i n_j$ has $2 \times 2 = 4$ components.  Because $\hat{\bm{n}}$ is a unit vector, we have the constraint $(\partial_i n_j)n_j = 0$.  That equation is actually 2 constraints, for $i=1$ and $2$.  Hence, the tensor $\partial_i n_j$ should have $4-2=2$ degrees of freedom.  The first leg of this tensor might have components parallel or perpendicular to $\hat{\bm{n}}$, but the second leg must be perpendicular to $\hat{\bm{n}}$.

We break $\partial_i n_j$ into parts where the first leg is parallel or perpendicular to $\hat{\bm{n}}$,
\begin{equation}
\partial_i n_j = -n_i B_j + \alpha_{ij},
\end{equation}
where $\bm{B}$ is a vector perpendicular to $\hat{\bm{n}}$ and $\alpha_{ij}$ is a tensor perpendicular to $\hat{\bm{n}}$.  Contracting both sides of this equation with $n_i$ gives
\begin{equation}
\bm{B}=-(\hat{\bm{n}}\cdot\bm{\nabla})\hat{\bm{n}}.
\end{equation}
Hence, $\bm{B}$ is the 2D version of the standard bend vector.  Because it is perpendicular to $\hat{\bm{n}}$ in 2D, it has one degree of freedom.  
Because it is invariant under the transformation $\hat{\bm{n}}\to-\hat{\bm{n}}$, it is a physical object that does not depend on this arbitrary choice of sign.

Now we are left with the tensor $\alpha_{ij}$ perpendicular to $\hat{\bm{n}}$.  In 2D, this tensor can be written as $\alpha_{ij}=S(\delta_{ij}-n_i n_j)$ for some scalar $S$.  Hence, the director gradient tensor becomes
\begin{equation}
\partial_i n_j = -n_i B_j + S(\delta_{ij}-n_i n_j),
\label{decomposition2D}
\end{equation}
Taking the trace of both sides of this equation gives
\begin{equation}
S=\bm{\nabla}\cdot\hat{\bm{n}}.
\end{equation}
Hence, $S$ is the 2D version of the standard splay scalar.  Because it is a scalar, it has one degree of freedom.  Just as in 3D, $S$ changes sign under the transformation $\hat{\bm{n}}\to-\hat{\bm{n}}$.  If we want a physical object that does not depend on the arbitrary sign of $\hat{\bm{n}}$, we can construct the splay vector $\bm{S}=S\bm{\hat{n}}=\bm{\hat{n}}(\bm{\nabla}\cdot\hat{\bm{n}})$.

Equation~(\ref{decomposition2D}) decomposes the director gradient tensor $\partial_i n_j$ into the two normal modes $\bm{B}$ and $S$ in 2D.  These two modes account for the two degrees of freedom in $\partial_i n_j$.  In 2D, there is no analogue for the twist $T$ or the biaxial splay $\bm{\Delta}$, or for saddle-splay.

\bibliography{saddlesplay2}

\end{document}